\documentclass[a4paper,12pt]{article}
\usepackage{graphicx} % Required for inserting images
\usepackage{amsmath, amsthm}
 \usepackage{array}
 \usepackage{float}
\usepackage{amssymb}
\usepackage{amsfonts}
\usepackage{dsfont}
\usepackage{booktabs}
\usepackage{graphicx}
\usepackage{multirow}
\usepackage[table, svgnames, dvipsnames]{xcolor}
\usepackage{natbib}
\usepackage[english]{babel}
\usepackage[nottoc]{tocbibind}
\usepackage{rotating}
\usepackage{pifont}% http://ctan.org/pkg/pifont
\usepackage{gensymb}
\usepackage{hyperref}
\usepackage{afterpage}
\usepackage{setspace}
\usepackage[font=small]{caption}

\usepackage{algorithm}
\usepackage{algpseudocode}

\usepackage{geometry}
\DeclareMathOperator*{\argmin}{{arg\,min}}

    \def \mY {\text{\boldmath$Y$}}

\def \varthetavec     {\text{\boldmath$\vartheta$}}

\title{Boosting Distributional Copula Regression for Bivariate
Binary, Discrete and Mixed Responses}
\author{Guillermo Brise\~no Sanchez, Nadja Klein, Hannah Klinkhammer, \\ Andreas Mayr}
\date{}

\begin{document}

\maketitle

\thispagestyle{empty} 
\begin{abstract}
\noindent  Motivated by challenges in the analysis of biomedical data and observational studies, we develop statistical boosting for the general class of bivariate distributional copula regression with arbitrary marginal distributions, which is suited to model binary, count, continuous or mixed outcomes. In our framework, the joint distribution of arbitrary, bivariate responses  is modelled through a parametric copula. To arrive at a model for the entire conditional distribution, not only the marginal distribution parameters but also the copula parameters are related to covariates through additive predictors. We suggest efficient and scalable estimation by means of an adapted component-wise gradient boosting algorithm with statistical models as base-learners. A key benefit of boosting as opposed to classical likelihood or Bayesian estimation is the implicit data-driven variable selection mechanism as well as shrinkage without additional input or assumptions from the analyst. To the best of our knowledge, our implementation is the only one that combines a wide range of covariate effects, marginal distributions, copula functions, and implicit data-driven variable selection. We showcase the versatility of our approach on data from genetic epidemiology, healthcare utilization and childhood undernutrition. Our developments are implemented in the \texttt{R} package \texttt{gamboostLSS}, fostering transparent and reproducible research. \\
\end{abstract}


\pagebreak
\setcounter{page}{1}
\onehalfspacing 
\section{Introduction}

Distributional regression models have gained considerable prominence  in statistical research over the last decade, thereby moving the focus from modelling the conditional mean of the response variable (as done in classical regression) towards modelling the entire conditional distribution. Having one model that describes the complete distribution is also of high relevance in biomedical research, since it allows to derive and understand relevant quantities such as variance or quantiles of biomarkers,  phenotypes or other outcomes of interest. Common examples are the construction of reference curves or growth charts, where skewness is often covariate-specific \citep[see e.g.,][]{IntPohAhrPig2016,GAMLSSREFERENCECURVE2}; or bivariate time-to-event data \citep{MarraRadiceJASACopulaSurvival}. 

Several distinct approaches to distributional regression for univariate responses exist \citep[see][for a recent review]{Kle2024}. The model class of our paper builds on generalised additive models for location, scale shape \citep[GAMLSS;][]{GAMLSSPAPERORIGINAL}, which allow to relate all distribution parameters of an arbitrary univariate parametric distribution to  covariates. While originally proposed for univariate responses, GAMLSS have been extended to accommodate regression models for multivariate responses \citep{Klein2014MultivariateDistReg}, although practically most existing approaches are limited to the bivariate case \citep[e.g.,][]{CraSab2012,VGAMBOOK, KleinKneibBIVCOPULASBAYES}. While parametric bivariate distributions such as the bivariate Gaussian, bivariate Bernoulli or bivariate Poisson offer an avenue for modelling bivariate responses, they also impose limitations on the distribution of the margins e.g.\ being univariate Gaussian or Poisson. A flexible alternative way to construct bivariate distributions are copulas \citep{Nelsen2006}. This approach allows to link arbitrary marginal distributions through a copula function, reflecting the association between the components. The literature on copula modelling is vast \citep[see e.g.,][for a review]{Smi2011}. 

 Reflecting the diversity of response types in our biomedical applications, in this paper we are particularly concerned with situations where the response variable is a bivariate vector $\boldsymbol{Y} = (Y_1, Y_2)^\top$ with components on possibly different domains that are expected to be associated with each other.  We hence focus on bivariate distributions constructed via  copulas and estimate all model-related quantities simultaneously instead of relying on a multi-step procedure. Recent contributions that employ a modeling and simultaneous estimation paradigm akin to ours can be found in \cite{MarRad2017} featuring a bivariate continuous response, \cite{MarraBIVBINARY} using bivariate binary outcomes, \cite{VanderwurpCountData} studying bivariate count responses, as well as \cite{KleinBinaryContinuous} analysing a mixed binary \& continuous response. All these contributions showed how to construct highly flexible bivariate copula regression models that are able to  accommodate a wide range of covariate effects as well as response types. Moreover, the substantial flexibility inherent in this model class of distributional copula regression models notably exacerbates the issue of variable selection -- a challenge that currently remains unaddressed within the specific models we are considering.

\begin{figure}[!t]
    \centering
    \includegraphics[scale=0.475]{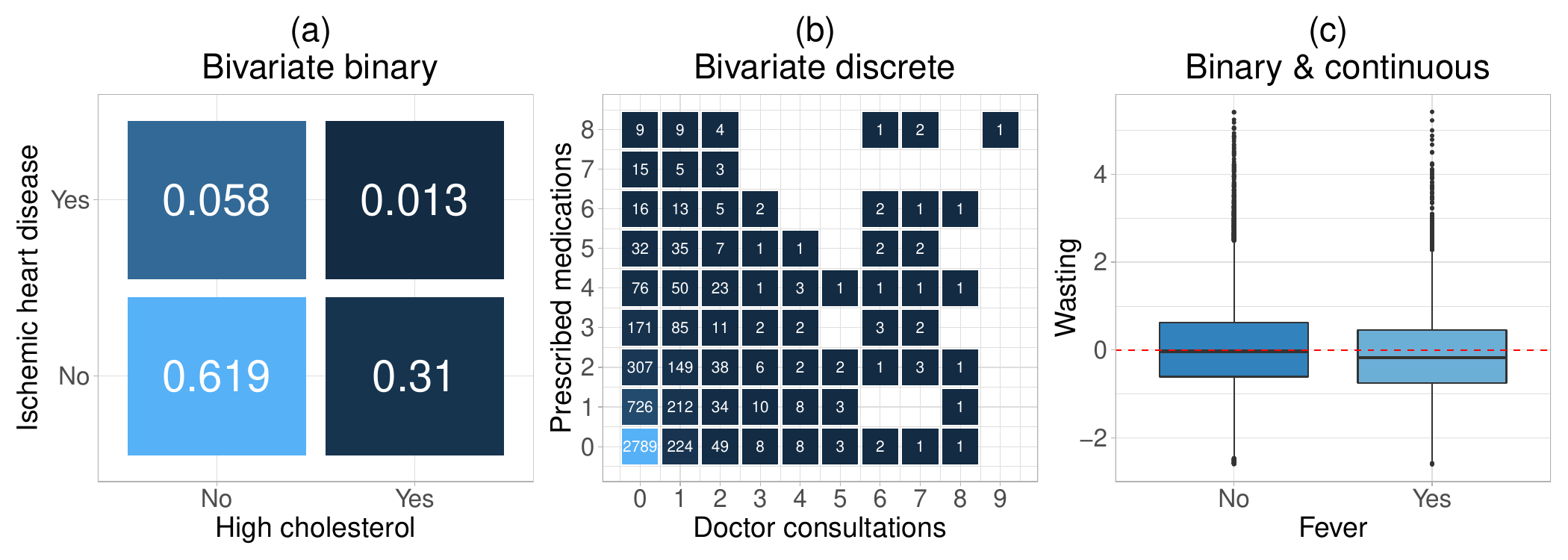}
    \caption{\small Responses in our applications analyzed in Section~\ref{SECTIONCaseStudy}.  (a) binary-binary response (numbers indicate proportions) in chronic ischemic heart disease; (b) count-count response (numbers indicate cases) in doctor visits and medical prescriptions; and (c) binary-continuous response in infant malnutrition in India. }\label{FIGUREThreeScatterplots}
\end{figure}

Our methodological contribution builds on the recent work by \cite{HansBoostDistReg} who previously integrated bivariate distributional copula regression models in the component-wise gradient boosting framework. This enables the estimation of all model-related quantities as well as  conducting variable selection in a data-driven manner instead of relying on significance-based heuristics or information criteria. However, one of the limitations of the approach by \cite{HansBoostDistReg} is that the response variables are both restricted to be strictly continuous. In many biomedical applications (but not only there), data is often recorded at a discretised scale (e.g.\ symptoms present yes/no) or the responses of interest actually depict a phenomenon expressed through discrete numbers/positive integers as in, for example, the number of doctor appointments and the number of prescription medications designated to a patient. At the time of writing a search in PubMed (\url{https://pubmed.ncbi.nlm.nih.gov}) returns $395{,}078$ and $24{,}439$ results for ``logistic regression'' and ``Poisson regression'' since 2010, respectively, highlighting the prevalence of this type of responses. It may also be the case, that the biomedical outcome is expressed as a combination of responses that lie in different domains, for example a binary indicator and a continuous measurement reflecting a disease (or symptom) indicator and an undernutrition score. These three aforementioned examples are the ones we consider later in Section~\ref{SECTIONCaseStudy} and the marginal distributions are visualized in Figure~\ref{FIGUREThreeScatterplots}. 

Recent work by \cite{StroemerDistReg} combined multivariate distributional regression with gradient-boosting in order to fit interpretable and highly flexible regression models in high-dimensional biomedical settings for bivariate continuous, bivariate binary and bivariate count responses. Their work considered two bivariate discrete distributions: the bivariate Poisson and the bivariate Bernoulli, which suffer from some limitations. On the one hand, the bivariate Poisson distribution is only able to model positive association structures between the margins though a ``covariance'' parameter. On the other hand, the bivariate Bernoulli distribution models the association between the marginal responses by means of the ``odds ratio'', whose ease of interpretation remains at best questioned, see for example \cite{OddsRATIOManual}. Furthermore, the marginal distributions of the components in the response vector are assumed to be of the same type, i.e.~the margins of a bivariate Poisson distribution must be univariate Poisson distributions. Such a restrictive assumption might not always be supported by the data. Therefore, the approach of \citet{StroemerDistReg} would be inappropriate for biomedical applications where the outcome of interest is composed of two responses emanating from different domains. For example, in one of our applications where we study childhood malnutrition via the joint distribution of \textit{wasting}, a continuous indicator for acute malnutrition as reflected by low weight for height (in comparison to a reference population), and a binary indicator for fever within the two weeks preceding a survey interview. 

The aim of this manuscript is threefold: First, we build upon \cite{HansBoostDistReg} and extend the class of boosting bivariate distributional copula regression models to arbitrary margins on different domains. Second, we expand the catalogue of copula functions and families of marginal distributions available for the publicly available \texttt{R} package \texttt{gamboostLSS} \citep{HofnerGAMBOOSTLSS}. These new additions allow for conducting data-driven variable selection and shrinkage in both low and high-dimensional applications, where the number of candidate variables ($p$) may greatly exceed the number of observations ($n$).  This can be applied to a wide range of data, and it also fosters transparent and reproducible research.  Third, we demonstrate the versatility and wide applicability of our approach through three diverse biomedical applications.

The rest of manuscript is structured as follows: Section~\ref{SECTIONDistCopReg} reviews the distributional copula regression for different types of responses and outlines our boosting algorithm. Section  \ref{SECTIONSimStudy} describes our simulation studies as well as their respective results. Section  \ref{SECTIONCaseStudy} presents the three case studies where we analyze data from epidemiological applications, in particular genetic epidemiology, healthcare and public health policy related to infant's malnutrition. We additionally illustrate the model-building process that involves selecting marginal distributions as well as copula distributions. Lastly, a discussion is given in Section~ \ref{SECTIONDiscussion}.

\section{Bivariate Distributional Copula Regression }\label{SECTIONDistCopReg}

\subsection{Model structure}

Distributional regression is a statistical framework that models the entire conditional response distribution using the covariate information at hand \citep{Kle2024}. Within the framework of GAMLSS, one assumes that the $i$-th observation of a response, with $i=1, \dots,n$, follows a parametric distribution with cumulative distribution function (CDF) $F$ and corresponding density $f$. In the context of bivariate responses considered here, the joint distribution of the random vector $\mY=(Y_{1i}, Y_{2i})^\top$ is denoted by
$
    P(Y_{1} \leq y_{1i},\ Y_{2} \leq y_{2i} \mid \boldsymbol{\vartheta}_i) = F_{1,2}(y_{1i}, y_{2i} \mid \boldsymbol{\vartheta}_i), 
$
where $F(\cdot\mid\varthetavec_i)$ represents the joint CDF parameterized through a $K$-dimensional parameter vector $\boldsymbol{\vartheta}_i=(\vartheta_{i1},\ldots,\vartheta_{iK})^\top$.  Rather than assuming a joint parametric distribution for $\mY$, we resort to a copula based approach using Sklar's theorem \citep{Nelsen2006}. This theorem states that any bivariate distribution can be written as  
\begin{align}\label{CopulaIsJointDistribution}
    F(y_{1i}, y_{2i} ; \boldsymbol{\vartheta}_i) =  C\left(F_1\left(y_{1i} \ ; \ \boldsymbol{\vartheta}^{(1)}_{i} \right),\ F_2\left(y_{2i}\; \ \boldsymbol{\vartheta}^{(2)}_{i} \right); \ \boldsymbol{\vartheta}^{(c)}_i \right),
\end{align}
where $C(\cdot, \cdot): [0,1]^2 \rightarrow [0,1]$ is the CDF of a bivariate parametric copula function with parameters $\boldsymbol{\vartheta}^{(c)}_i$. The copula \emph{links} the possibly different parametric marginal distributions with CDFs $F_1,F_2$ and respective parameter vectors $\boldsymbol{\vartheta}^{(1)}_{i} $, $\boldsymbol{\vartheta}^{(2)}_{i}$.  In what follows, we consider one-parametric bivariate copulas, and refer to $\boldsymbol{\vartheta}^{(c)}_i = \vartheta^{(c)}_i$ as the corresponding scalar association parameter that determines the strength of the association between the marginal responses. Table~\ref{CopulaTables}  details  the implemented copulas in the \textsf{R} add-on package \textsf{gamboostLSS}.
\begin{sidewaystable}[p]%[htbp]
    \centering\renewcommand{\arraystretch}{2.5}\small 
    \begin{tabular}{ccccc}
    \toprule
Copula & $C(F_1, F_2; \vartheta^{(c)})$ & Range of $\vartheta^{(c)}$   & Link &   Kendall's $\tau$  \\
    \midrule
     Gauss   & $\Phi_2 (\Phi_1^{-1}(F_1), \Phi_1^{-1}(F_2); \vartheta^{(c)} )$ & $\vartheta^{(c)} \in [-1, 1]$  & $\tanh^{-1}(\vartheta^{(c)})$ & $\frac{2}{\pi} \arcsin(\vartheta^{(c)})$ \\
    Clayton &  $(F_1^{-\vartheta^{(c)}} + F_2^{-\vartheta^{(c)}} - 1)^{-1/\vartheta^{(c)}}$ & $\vartheta^{(c)} \in (0, \infty)$ & $\log(\vartheta^{(c)})$ & $\frac{\vartheta^{(c)}}{\vartheta^{(c)} + 2}$ \\
    Gumbel  & $\exp\left[ - \left\{ (-\log(F_1))^{\vartheta^{(c)}} + (-\log(F_2))^{\vartheta^{(c)}} \right\}^{\tfrac{1}{\vartheta^{(c)}}} \right]$  & $\vartheta^{(c)} \in [1, \infty)$ & $\log(\vartheta^{(c)} - 1)$ & $1 - \frac{1}{\vartheta^{(c)} }$\\
     Frank   & $-{\vartheta^{(c)}}^{-1} \log\Big(1 + (\exp(-\vartheta^{(c)} F_1) - 1 ) \cdot$ & $\vartheta^{(c)} \in \mathbb{R} \setminus \{0\} $ & $\vartheta^{(c)}$ & $ 1- \frac{4}{\vartheta^{(c)}}[ 1- D_1(\vartheta^{(c)})] $ \\
     & $(\exp(-\vartheta^{(c)} F_2) - 1 ) / (\exp(-\vartheta^{(c)}) - 1) \Big)$ & & &  \\ 
     AMH  & $F_1 F_2 / (1 - \vartheta^{(c)} (1 - F_1) (1 - F_2))$ & $\vartheta^{(c)} \in [-1, 1]$ & $\tanh^{-1}(\vartheta^{(c)})$  & $1 -\frac{2}{3}  {\vartheta^{(c)}}^{2} (\vartheta^{(c)} + (1-\vartheta^{(c)})^2 \log(1 - \vartheta^{(c)} ) ) $  \\
      FGM  & $F_1 F_2 / (1 + \vartheta^{(c)} (1 - F_1) (1 - F_2))$  & $\vartheta^{(c)} \in [-1, 1]$ & $\tanh^{-1}(\vartheta^{(c)})$  & $\frac{2}{9} \vartheta^{(c)}$ \\
 Joe & $1 - ((1 - F_1)^{\vartheta^{(c)} } + (1 - F_2)^{ \vartheta^{(c)}} - (1 - F_1)^{ \vartheta^{(c)} }  (1 - F_2)^{ \vartheta^{(c)} })^{ ( 1/\vartheta^{(c)}  ) } $  & $\vartheta^{(c)} \in [1, \infty)$ &  $\log(\vartheta^{(c)} - 1)$  & $1 + \frac{4}{ {\vartheta^{(c)} }^{2} } \int_{0}^{1} x \log(x) (1 - x)^{2 ( 1 - \vartheta^{(c)} ) / \vartheta^{(c)} } dx $ \\
         \bottomrule
    \end{tabular}
    \caption{\small Details of implemented copulas. The functions $\Phi_1^{-1}(\cdot)$ and $\Phi_2(\cdot)$ denote the quantile function and CDF of the univariate and bivariate standard normal distributions, respectively.  Rotated copulas by 90, 180 and 270 degrees are respectively defined as: $C_{90} = F_2 - C(1-F_1, F_2; \vartheta^{(c)})$, $C_{180} = F_1 + F_2 - 1 + C(1-F_1, 1-F_2; \vartheta^{(c)})$ and $C_{270} = F_1 - C(F_1, 1-F_2; \vartheta^{(c)})$. 
    The term $D_1(\vartheta^{(c)}) = \int_0^{\vartheta^{(c)}} \frac{t}{\exp(t) -1 }dt$ is the Debye function and $\Phi_2$ denotes the CDF of the bivariate Gaussian distribution with correlation coefficient $\vartheta^{(c)}$. Finally, AMH stands for Ali-Mikhail-Haq and FGM stands for Farlie-Gumbel-Morgenstern.}
    \label{CopulaTables}
\end{sidewaystable}

Let now $K=K_1+K_2+K_c=K_1+K_2+1$ denote the total number of distribution parameters in the bivariate distribution and  $\boldsymbol{\vartheta}^{(1)}_{i}=(\vartheta_{i1}^{(1)},\ldots,\vartheta_{iK_1}^{(1)})^\top$,  $\boldsymbol{\vartheta}^{(2)}_{i}=(\vartheta_{i1}^{(2)},\ldots,\vartheta_{iK_2}^{(2)})^\top$, be the vectors containing all parameters that correspond to the respective marginal distributions. All $K$ parameters of the bivariate distribution are then stored in the vector
 $\boldsymbol{\vartheta}_i = \left((\boldsymbol{\vartheta}^{(1)}_{i})^\top, (\boldsymbol{\vartheta}^{(2)}_{i})^\top, \boldsymbol{\vartheta}^{(c)}_{i}\right)^\top$.
 The distributional copula regression approach allows each component of $\varthetavec_i$  to depend on the vector of covariates denoted by $\boldsymbol{x}_i$ by means of  structured additive predictors and suitable link functions $g(\cdot)$ with corresponding inverse or response functions $h(\cdot)\equiv g^{-1}(\cdot)$ that ensure potential parameter space restrictions, that is,
\begin{align}\label{DistRegApproachPredictor}
g_{k}^{(\bullet)}\left( \vartheta_{ik}^{(\bullet)} \right) =\eta_{ik}^{(\bullet)}  &= \beta_{0k}^{(\bullet)}  +  \sum_{r = 1}^{ P_{k}^{(\bullet)}} s_{ rk }^{(\bullet)} ( x_{ir}).
\end{align}
The symbol $\bullet\in\lbrace 1, 2, c\rbrace$. The summation limit $P_{k_{} }^{(\bullet)}$ emphasizes that the individual parameters $\vartheta_{ik_{}  }^{(\bullet)}$ do not necessarily have to be modelled using the same subset of covariates. The coefficients $\beta_{0 k_{} }^{(\bullet)}$ are parameter-specific intercepts and $s_{ rk }^{(\bullet)} (\cdot)$ are smooth functions that can accommodate a wide range of functional forms of the covariates, such as linear, non-linear, or spatial effects. Each covariate effect is modelled by appropriate basis function expansions of the form: 
$$
s_{ rk }^{(\bullet)} (x)=\sum_{l=1}^{L_{rk}^{(\bullet)}}\beta_{rk,l}^{(\bullet)}B_{rk,l}^{(\bullet)}(x),
$$ 
where $B_{rk,l}^{(\bullet)}(x)$ is a suitable basis function evaluated at the observed covariate value and $\beta_{rk,l}^{(\bullet)}$ are generic coefficients to be estimated.  As mentioned earlier and further emphasized by the summation index $P^{(\bullet)}_k$ shown in Equation~\eqref{DistRegApproachPredictor}, there may not be strong a-priori evidence of which subset of covariates (or if any at all) has an effect on the individual parameters $\vartheta_{k}^{(\bullet)}$ of the bivariate distribution $F(\cdot,\cdot;\varthetavec)$. Therefore, we resort to component-wise gradient-boosting or statistical boosting \citep{EvolutionOFBOOSTINGALGORITHMS} to estimate all coefficients simultaneously, see Section~\ref{BoostingSection} for a description of the estimation algorithm. Compared to \citet{HansBoostDistReg}, who assumed that $\mY$ has two continuous components, our approach can also handle binary, discrete, or a combination of binary and continuous components $Y_{ji}$, $j=1,2$, that make up the bivariate response $\mY$.  As a final remark, Sklar's theorem guarantees that the copula characterising the joint distribution of $\mathbf{Y}_i$ is unique only if the marginal responses are continuous. 
However, since we are dealing with a regression model structure, i.e.\ the model is cast in terms of the coefficients $\beta_{rk,l}^{(\bullet)}$ instead of being parameterised directly via the parameters $\vartheta_k^{(\bullet)}$, this issue is diminished \citep[cf., e.g.,][]{Trivedi2017}.

\subsection{Relevant examples of bivariate responses}
In the following, we describe the  bivariate response types relevant for our applications. The respective choices of corresponding marginal distributions are summarized in Table~\ref{TableMarginalDistributions} together with main characteristics, such as expectation and variance.
\paragraph{Bivariate binary responses}
We begin by considering the case $Y_{ji} \in \{0,1\}$, $j = 1,2$. The individual marginal probabilities of observing $y_{ji} = 1$ are modelled via $P\left( Y_{ji} = 1;\vartheta_i^{(j)} \right) = \vartheta_i^{(j)} = h^{(j)}\left( \eta_{i}^{(j)} \right) =: p^{1(j)}_{i}$, $j = 1,2$, where the response function can be any function suitable for parameters whose range is the unit interval $[0,1]$, e.g.\ logit, probit and cloglog link functions. The joint probability of both $y_{1i}$ and $y_{2i}$ being equal to one is obtained using
$$
P(Y_{1i} = 1, Y_{2i} = 1;\varthetavec_i) = C\left(P\left(Y_{1i} = 1; \vartheta_i^{(1)} \right), P\left(Y_{2i} = 1; \vartheta_i^{(2)} \right) ;  \vartheta^{(c)}_{i} \right) =: p^{11}_{i}.
$$
The joint probability mass function consists of the four possible outcomes of the binary responses, that is $(y_{1i}, y_{2i}) \in \{(1,1), (1,0), (0,1), (0,0)\}$. This leads to the following log-likelihood contribution of the $i$-th observation:
\begin{equation}\label{BivBinaryLogLik}\begin{aligned}
    \ell_i &= y_{1i} y_{2i} \log\left(p^{11}_{i} \right) +  y_{1i} (1-y_{2i}) \log\left( p^{1(1)}_{i} - p^{11}_{i} \right) + (1-y_{1i}) y_{2i} \log\left( p^{1(2)}_{i} - p^{11}_{i} \right) \\ 
    &\quad + (1-y_{1i}) (1-y_{2i}) \log\left(1 - p^{1(1)}_{i} - p^{1(2)}_{i} + p^{11}_{i} \right) .
\end{aligned}\end{equation}
Note that our implementation allows the individual marginal probabilities to be modelled using identical or different link functions, for example margin~1 using the probit link and margin~2 using the logit link.

\paragraph{Bivariate discrete responses}
Each marginal response is a count variable, that is\ $Y_{ji} \in \mathbb{N}_{\geq 0}$, $j = 1,2$. Here we denote with $P\left( Y_{ji} \leq y_{ji} ; \boldsymbol{\vartheta}_i^{(j)} \right) = F_{j}\left(y_{ji}; \boldsymbol{\vartheta}_i^{(j)} \right)$ the marginal CDFs, and with $P\left(Y_{ji} = y_{ji} ;\boldsymbol{\vartheta}_i^{(j)} \right) = f_{j}\left( y_{ji} ; \boldsymbol{\vartheta}_i^{(j)} \right)$ the marginal PDFs of $Y_{ji}$. Similar to \cite{VanderwurpCountData}, we compute $P\left(Y_{ji} = y_{ji} - 1 ; \boldsymbol{\vartheta}_i^{(j)} \right) = F_{j} \left(y_{ji} ; \boldsymbol{\vartheta}_i^{(j)} \right)-f_{j}\left(y_{ji} ; \boldsymbol{\vartheta}_i^{(j)} \right)$ in order to avoid a (trivial) evaluation of the CDF of $Y_{ij}$ with a negative argument in case that $y_{ij} = 0$, $j = 1, 2$. The log-likelihood function of the $i$-th observation is then given by:
{\footnotesize
\begin{align}\label{BivDiscreteLoglik}
    \ell_i &= \log\Big( C(F_1(y_{1i};\boldsymbol{\vartheta}_i^{(1)}), F_2(y_{2i};\boldsymbol{\vartheta}_i^{(2)});\vartheta_i^{(c)} ) -  C(F_1(y_{1i};\boldsymbol{\vartheta}_i^{(1)}) - f_1(y_{1i};\boldsymbol{\vartheta}_i^{(1)}), F_2( y_{2i};\boldsymbol{\vartheta}_i^{(2)};\vartheta_i^{(c)}) ) \nonumber\\
    &\phantom{====} 
    -C(F_1(y_{1i};\boldsymbol{\vartheta}_i^{(1)}), F_{2} (y_{2i};\boldsymbol{\vartheta}_i^{(2)})-f_{2}(y_{2i};\boldsymbol{\vartheta}_i^{(2)});\vartheta_i^{c} ) 
    \nonumber\\
    &\phantom{====}+ C(F_1( y_{1i};\boldsymbol{\vartheta}_i^{(1)} ) - f_1(y_{1i};\boldsymbol{\vartheta}_i^{(1)}), F_{2} (y_{2i};\boldsymbol{\vartheta}_i^{(2)})-f_{2} (y_{2i};\boldsymbol{\vartheta}_i^{(2)});\vartheta_i^{c} ) \Big).
\end{align} }
We have implemented various discrete distributions, including common ones such as the Poisson and Geometric distributions. Additionally, we have integrated two-parameter count distributions designed for over-dispersed data such as the Negative Binomial (Type I).  Furthermore, to handle count data characterized by an excess of zero observations , we have included zero-inflated and zero-altered distributions. These include models like the  Zero-Altered Logarithmic, Zero-Altered Negative Binomial, Zero-inflated Poisson and Zero-Inflated Negative Binomial distributions. For more specifics on the  parameterizations of these distributions, please refer to  \cite{GamlssdistributionsBook}.

\begin{sidewaystable}[p]
 \centering\small
    \begin{tabular}{ l c c l l c }
    \toprule
      Distribution  & $E(Y)$  & $Var(Y)$ & Parameters \& range & Links & Response range  \\
         \midrule
      \multirow{3}{*}{Bernoulli}  & \multirow{3}{*}{ $\vartheta_{1}$ } & \multirow{3}{*}{ $\vartheta_{1} \left(1 - \vartheta_{1} \right) $ } & \multirow{3}{*}{$\vartheta_{1} \in [0,1] $ } & logit & \multirow{3}{*}{$y \in \{0,1\}$ }\\
      &  & & & probit \\ 
      &  & & & cloglog \\
     \\
      Gaussian  & $\vartheta_{1}$  & $\vartheta_{2}^2$  & $\vartheta_{1} \in \mathbb{R}, \vartheta_{2} > 0  $ & Identity, log & $y \in \mathbb{R}$   \\
     \\
    Poisson & $\vartheta_{1}$ & $\vartheta_{1}$ & $ \vartheta_{1} > 0 $ & log &  $y \in \mathbb{N}_+$ \\
    \\
       Geometric  & $\vartheta_{1}$ & $\vartheta_{1} + \vartheta_{1}^2$ & $\vartheta_{1} > 0  $  & log & $y \in \mathbb{N}_+$ \\
       \\
       Negative Binomial (I)  & $\vartheta_{1}$ & $\vartheta_{1} + \vartheta_{2}\vartheta^2_{1}$ & $\vartheta_{2}, \ \vartheta_{2} > 0$  & log, log & $y \in \mathbb{N}_+$\\
       \\
       Zero-Altered Logarithmic & $ \frac{(1 - \vartheta_{2}) \alpha \vartheta_{1} }{ (1 - \vartheta_{1}) } $ &
       $ \frac{ (1 - \vartheta_{2}) \alpha \vartheta_{1} (1 - (1 - \vartheta_{2}) \alpha \vartheta_{1} ) }{  \left( 1- \vartheta_{1} \right)^2 } $ & $\vartheta_{1}, \ \vartheta_{2} \in [0,1] $  & logit, logit & $y \in \mathbb{N}_+$\\
       \\
       Zero-Inflated Poisson & $(1 - \vartheta_{2})\vartheta_{1}$ & $\vartheta_{1} (1 - \vartheta_{2}) ( 1 + \vartheta_{1} \vartheta_{2})$ & $\vartheta_{1}, \ \vartheta_{2} \in [0,1]  $ & log, logit & $y \in \mathbb{N}_+$\\
       \\
       Zero-Altered Negative Binomial & $c \vartheta_{1}$ & $c \vartheta_{1} + c\vartheta_{1}^2 (1 + \vartheta_{2} - c)$ & $\vartheta_{1}, \vartheta_{2} > 0, \ \vartheta_{3} \in [0,1]  $ & log, log, logit & $y \in \mathbb{N}_+$  \\
       \\
       Zero-Inflated Negative Binomial & $(1 - \vartheta_{3}) \vartheta_{1}$ & $\vartheta_{1} (1 - \vartheta_{3}) + \vartheta_{1}^2 (1 - \vartheta_{3})(\vartheta_{2} + \vartheta_{3})  $  & $\vartheta_{1}, \vartheta_{2} > 0, \ \vartheta_{3} \in [0,1]  $ & log, log, logit & $y \in \mathbb{N}_+$\\
         \bottomrule
    \end{tabular}
    \caption{\small Details of newly implemented univariate marginal distributions for binary, continuous and discrete responses to be used together with copulas in the \texttt{gamboostLSS} package. For the Zero-Altered Logarithmic distribution, the term $\alpha = - [ \log(1 - \vartheta_{1})]^{-1} $. For the Zero-Altered Negative Binomial distribution, the term $c = \frac{1 - \vartheta_{3}}{\left( 1 - (1 + \vartheta_{1}\vartheta_{2})^{-1/\vartheta_{2} } \right) }$. All distributions use the parameterisation from \cite{GamlssdistributionsBook}.  }
    \label{TableMarginalDistributions}    
\end{sidewaystable}

\paragraph{Bivariate mixed binary-continuous responses}
When one response component is continuous and the other binary, we follow \cite{KleinBinaryContinuous} and resort to a latent variable representation of the regression model for the binary component. Without loss of generality, let the first component of the bivariate vector be the binary variable, that is, $Y_{1i} \in \{0,1\}$. The binary response $Y_{1i}$ is then determined by an unobserved, latent variable $Y^{*}_{1i}$ with parametric CDF $F^*_1(y^*_{1i};\varthetavec_i^{(1)})$ through the mechanism: $Y_{1i} = \mathds{1}(Y^*_{1i} > 0)$, where $\mathds{1}(\cdot)$ is the indicator function. Then it follows that
 $P(Y_{1i} = 0;\vartheta_i^{(1)}) = F_1(0;\vartheta_i^{(1)}) = F^{*}_1(0;\vartheta_i^{(1)}) = P(Y^{*}_{1i} \leq 0;\vartheta_i^{(1)})$, in other words, the CDFs of the binary and latent variables coincide at $y_{1i} = y^*_{1i} = 0$. With this representation, the joint bivariate distribution can be written as:
 \begin{align*}
     P(Y_{1i} = 0, Y_{2i} \leq y_{2i};\varthetavec_i) = P(Y^*_{1i} \leq 0, Y_{2i} \leq y_{2i}) = C(F_{1}^*(0;\vartheta_i^{(1)}), F_2(y_{2i};\varthetavec_i^{(2)});\vartheta_i^{(c)}),
 \end{align*}
from which we obtain the log-likelihood contribution:
{\footnotesize
\begin{align}\label{BinContLoglik}
     \ell_i &= (1 - y_{1i}) \log\left( \frac{\partial C(F_{1}(0;\vartheta_i^{(1)}), F_{2}(y_{2i};\varthetavec_i^{(2)});\vartheta_i^{(c)} ) }{\partial F_2(y_{2i};\varthetavec_i^{(2)}) }  \right) +  y_{1i} \log\left(1 -  \frac{\partial C(F_{1}(0;\vartheta_i^{(1)}), F_{2}(y_{2i};\varthetavec_i^{(2)});\vartheta_i^{(c)} ) }{\partial F_2(y_{2i};\varthetavec_i^{(2)}) }  \right)\nonumber\\
     &\quad+ \log(f_2(y_{2i};\varthetavec_i^{(2)})). 
\end{align} }
The link function for the binary margin can be set to logit, probit or cloglog. We have also added the Gaussian distribution to the catalogue of distributions for continuous responses of our software implementation, since it aligns with previous analyses of univariate malnutrition indicators \citep[cf.][]{KleinBinaryContinuous} used in our third case study in Section~\ref{SectionApplicationMixedBinCont}.  

\subsection{Estimation via component-wise gradient boosting}\label{BoostingSection}
We propose to estimate all model coefficients simultaneously via a component-wise gradient boosting algorithm with regression type base-learners, also often referred to as model-based boosting or statistical boosting \citep{FriedmanSOLO2000BOOSTING, BuehlmannHothornBOOSTING}. While boosting is a general concept from machine learning, it has also been extended towards estimating statistical models   \cite{EvolutionOFBOOSTINGALGORITHMS}. The term \textit{component-wise} highlights that this particular boosting framework fits the base-learners (components) one-by-one and greedily updates the model by updating only the best-performing component \citep{mboost2point0}. More concretely, every smooth function of the covariates in Equation~\eqref{DistRegApproachPredictor} is cast using a base-learner denoted by $b_r^{(\bullet)}(x_{ir})$, $\bullet\in\lbrace 1, 2, c\rbrace$. Each base-learner is typically based on a single explanatory variable specific to the type of covariate effect. The base-learners can be for example linear functions (corresponding to linear effects of covariates), P-splines for non-linear effects or Gaussian Markov Random Fields for structured discrete spatial effects (in which case the covariate would be bivariate corresponding to spatial coordinates, rather than univariate). We refer to \cite{mboost2point0} and \cite{Mayr2012GAMBOOSTLSS} for a complete list of the currently implemented base-learners  and to \cite{mayr2023linear} for a recent approach to choose the most appropriate ones. 

Estimating the model coefficients corresponds to solving the optimization problem:
\begin{equation*}
    \boldsymbol{\hat{\eta} } = \argmin_{\eta} \Big[ E_{\boldsymbol{Y}} \left\{ \omega\left(\boldsymbol{Y};  \boldsymbol{\eta }  \right) \right\} \Big],
\end{equation*}
where the vector $\boldsymbol{\eta} = \left(\boldsymbol{\eta}^{(1)}, \boldsymbol{\eta}^{(2)}, \boldsymbol{\eta}^{(c)} \right) \in \mathbb{R}^K$ contains all additive predictors corresponding to the parameters of the bivariate distribution and  $\boldsymbol{\hat{\eta}}$ denotes their estimates. The term $\omega(\cdot)$ represents the loss function, which in our case corresponds to the negative log-likelihood of the regression model, that is, $\omega(\cdot) = - \ell_i(\cdot)$. In general, minimizing the expectation of the loss is intractable. In practice, given a sample of $i = 1, \dots, n$ observations, one minimises the \textit{empirical risk} $\frac{1}{n} \sum_{i=1}^{n} \omega(\boldsymbol{y}_i; \boldsymbol{\eta}_i)$ iteratively. In each boosting iteration, the algorithm fits each of the pre-specified base-learners of each distribution parameters individually to the negative gradient of the loss function w.r.t.\ to the additive predictors of the parameters (also sometimes referred to as \textit{pseudo-residuals)}, i.e.\ $- \partial \omega( \boldsymbol{y}_i; \boldsymbol{\eta}_i) / \partial \eta^{(\bullet)}_{ k }$. Only the best-fitting base-learner  is selected and a ``weak'' update of the model is conducted. Since the type of base-learner for each covariate is unmodified throughout the fitting process, the final effect of the covariate remains of the same type. The fitting procedure is run for a pre-specified number of iterations denoted by $\texttt{m}_{\texttt{stop}}$, which plays a similar role like the penalty parameter ``$\lambda$'' of the LASSO \citep{LASSOANDBOOSTINGRELATIONSHIPMSTOP}, and acts as the  main tuning parameter. In our case, we conduct non-cyclical updates \citep{Thomas2017NONCYLCIC}, which means that only one out of all additive predictors is updated per fitting iteration. Only the update which leads to highest decrease in the empirical risk is selected to be carried out. By conducting \textit{early stopping}, i.e.\ using $\texttt{m}^{\texttt{opt}}_{\texttt{stop}} < \texttt{m}_{\texttt{stop}}$ fitting iterations, some base-learners will effectively be left out of the model, since they were not selected in any iteration. Hence early stopping results in intrinsic, data-driven variable selection as well as shrinkage of  covariate effects. Algorithm~\ref{boostingAlgorithm} provides a detailed description of our adopted procedure for a generic distributional regression model including a mechanism for faster tuning of $\texttt{m}_{\texttt{stop}}$. 
\begin{algorithm}[p]
\caption{\small Non-cyclic boosting for distributional copula regression with faster tuning of fitting iterations $\texttt{m}_{\texttt{stop}}$ by means of out-of-bag (\textit{oobag}) risk.}\label{boostingAlgorithm}
\begin{algorithmic}
\Require \\
Define the base-learners $b^{(\bullet)}_{r}(x_r)$ for $r = 1, \dots, P_{vk}$, $\bullet = 1, 2, c$. 
\State Set the step-length $\texttt{s}_{\texttt{step}}  \ll 1$ as well as the (non-optimal) number of fitting iterations $\texttt{m}_{\texttt{stop}}$. 
\State Set weights indicating the training and $\texttt{m}_{\texttt{stop}}$-tuning partitions of the sample $n_{\text{train}}$, $n_{\texttt{mstop}}$.
\State Set type of stabilisation to be applied to the negative gradient vector ($L_2$, median absolute deviation or none).
\State (1) Initialise all predictors $\hat{\eta}^{(\bullet)}_{k}$ corresponding to $\vartheta_k^{(\bullet)} \in \boldsymbol{\vartheta}$ with offset values $\hat{\eta}^{(\bullet)}_{k, [0]}$. 
\For{$m = 1, \dots, \texttt{m}_{\texttt{stop}}$}
\For{$k = 1, \dots, K$ in $\vartheta_k^{(\bullet)} \in \boldsymbol{\vartheta}$ }
\State (a) Evaluate the parameter-specific negative gradient vector $\boldsymbol{-g}^{(\bullet)}_{k, [m]}$ 
    \begin{equation*}
\boldsymbol{-g}^{(\bullet)}_{k, [m]} =
\left( \boldsymbol{-g}^{(\bullet)}_{k, [m]} (\boldsymbol{x}_i)  \right)_{i = 1, \dots, n_{\text{train}}} = 
- \left( \left.\frac{ \partial \omega\left( \boldsymbol{y}_i, \boldsymbol{\hat{\eta}}_{i} \right)  }{ \partial \eta^{(\bullet)}_{k} }\right\vert_{ \boldsymbol{\hat{\eta} } = \boldsymbol{\hat{\eta} }_{[m-1] }(\boldsymbol{x}_i) }  \right)_{i = 1, \dots, n_{\text{train} } }.
    \end{equation*}
\State (b) Fit $\boldsymbol{-g}^{(\bullet)}_{k, [m]}$ to each parameter-specific base-learner $b^{(\bullet)}_{k,j}(x_j)$.
\State (c) Select the best-fitting base-learner $\hat{b}^{(\bullet)}_{k,j^{\star}}$ via residual sum of squares criterion.
\begin{equation*}
j^{\star} = \argmin_{j \in 1, \dots P_k^{(\bullet)} } \sum_{i = 1}^{ n_{ \text{train} } } \left( -g^{(\bullet)}_{k, [m]}(\mathbf{x}_i) -  \hat{b}^{(\bullet)}_{k,j^{}}(x_{i})  \right)^2.
\end{equation*}
\State (d) Compute loss reduction of a weak update using $\hat{b}^{(\bullet)}_{k,j^{\star}}$.
\begin{equation*}
 \Delta \omega_{\vartheta^{(\bullet)}_k} = \sum_{i = 1}^{ n_{ \text{train} } } \omega\left( \boldsymbol{y}_i; \hat{\eta}_k + \texttt{s}_{\texttt{step}}   \hat{b}^{(\bullet)}_{k,j^{\star}}( x_{ij^{\star}} ) \right).
\end{equation*}
\EndFor
\State (2) Update the parameter with highest loss reduction $\vartheta^{(\bullet)^\star }_{k} = \argmin_{ \vartheta^{(\bullet) }_{k} \in \boldsymbol{\vartheta} } \left( \Delta \omega_{\vartheta^{(\bullet)}_k} \right)$:
\begin{equation*}
    \hat{\eta}^{(\bullet)*}_{k, [m]}(\boldsymbol{x}_i) =
     \hat{\eta}^{(\bullet)*}_{k, [m-1]}(\boldsymbol{x}_i) + \texttt{s}_{\texttt{step}}  \cdot \hat{b}^{(\bullet)}_{k,j^{\star}}( x_{ij^{\star}} ).
\end{equation*}

\State (3) For the remaining parameters $\vartheta^{(\bullet)}_k \neq \vartheta^{(\bullet)^\star }_k$, set $\hat{\eta}^{(\bullet)}_{k, [m]} (\boldsymbol{x}_i) = \hat{\eta}^{(\bullet)}_{k, [m-1]} (\boldsymbol{x}_i) $.
\State (4) Compute the out-of-bag risk at iteration $[m]$ : 
\begin{equation*}
\text{risk}_{\text{oobag}, [m]} = \sum_{i = 1}^{n_{ \texttt{mstop} }}  \hat{\omega}\left( \left. \boldsymbol{y}_i; \boldsymbol{\hat{\eta_{i}}}  \right\vert_{\hat{\eta} = \hat{\eta}_{[m]} (\boldsymbol{x}_i) }   \right)  .
\end{equation*}
\EndFor
\State (5) Determine $\texttt{m}^{\texttt{opt}}_{\texttt{stop}}$ by means of the out-of-bag-risk: 
\begin{equation*}
     \texttt{m}^{ \texttt{opt} }_{\texttt{stop}} = \argmin_{m \in 1, \dots , \texttt{m}_{\texttt{stop}} } \text{risk}_{\text{oobag}, [m]}.
\end{equation*}

\end{algorithmic}
\end{algorithm}

\afterpage{\clearpage}
\section{Simulation Study}\label{SECTIONSimStudy}

In this section we summarise the main findings of  our simulation study. We consider three response scenarios in Sections~\ref{sec:s1}--\ref{sec:s3}, one for bivariate binary, count and mixed outcomes each. The main goals are to evaluate (i) estimation, (ii) variable selection and (iii) predictive performance of our proposed bivariate copula approach compared to the benchmark of estimating two separate (and thus independent) univariate models. The code used to reproduce the simulations can be found in the following repository: \url{https://github.com/GuilleBriseno/BoostDistCopReg_BinDiscMix}.

\paragraph{General settings} All boosting models were fitted using the \texttt{gamboostLSS} package.  A training data set of $n_{\text{train}} = 1000$ observations and a fixed step-length  of $\texttt{s}_{\texttt{step}} = 0.1$ for all distribution parameters are used. Following \cite{Mayr2012GAMBOOSTLSS} as well as \cite{HansBoostDistReg}, the stopping iteration $\texttt{m}_{\texttt{stop}}$ is optimised by minimising the out-of-bag empirical risk using a validation data set with $n_{\texttt{mstop}} = 1500$ observations emanating from the same underlying distribution (see step (4) in Algorithm~\ref{boostingAlgorithm}).
Similar to \cite{HansBoostDistReg}, we  apply $L_2$-stabilisation to the parameter-specific gradients in order to obtain similar step-lengths among the various dimensions of the model, see \cite{HofnerGAMBOOSTLSS} for details on gradient stabilisation. The performance of the copula and univariate models is evaluated using multivariate proper scoring rules (negative log-likelihood and energy score), both oriented such that lower values indicate better performance. The energy score is computed using the \texttt{scoringRules}  package. We include univariate distribution-specific evaluation criteria as well, although we remark that these criteria do not take the dependence between the responses into account. For binary responses, we use the Brier score and the area under the curve (AUC). For the remaining discrete and mixed responses we compute the univariate mean squared error of prediction (MSEP) comparing the true $Y_j$ with its prediction $\hat Y_j$, $j=1,2$. All of the aforementioned scores are summarized in Table~\ref{SimulationsAllMetrics} and computed as the average over a separate partition of the synthetic data consisting of $n_{\text{test}} = 1000$ observations that are not used in the fitting process or for tuning. The bivariate observations are are generated using the \texttt{VineCopula} package. Lastly, we report the selection rates of informative as well as non-informative covariates for both copula and benchmark models for each distribution parameter in Table~\ref{SimulationsSelectionRates}. Each scenario is run using 100 independent synthetic datasets. In the following, we present the main configurations specific to each response scenario along with the key findings. 

\subsection{Bivariate binary responses}\label{sec:s1}
\paragraph{Data generation} 
We consider three data generating processes (DGPs) with increasing number of noise variables. Specifically, we generate $p_1 = 10$, $p_2 = 100$ and lastly $p_3=1000$ covariates, of which only six  have are truly informative in one or several of the distribution parameters.  The bivariate distribution of the binary components is created using a Gaussian copula with varying correlation between the margins. On average, the dependence between the margins of the synthetic data in terms of Kendall's $\tau$ lies within $[-0.993; \ 0.993]$, i.e.\ it ranges between very strong negative to very strong positive dependence. We generate the first margin from a probit model and the second margin from a cloglog model. Thus, the model has $K=3$ distribution parameters and the DGP with $p_3$ covariates represents a high-dimensional setting with $p_3 \gg n$, resulting in effectively $p_3\times K=3000$ covariates since we fit all regressors to each distribution parameter. For this scenario we only consider DGPs with linear effects of the covariates, which reflect the data analysed in Section~\ref{SectionApplicationBivBinary}. Following \cite{StroemerDistReg}, we generate the $p_q$, $q=1,2,3$  covariates by sampling from a multivariate Gaussian distribution with Toeplitz covariance structure of the form $\Sigma_{ij} = \rho^{\mid i - j\mid}$ for $1 \leq i, j \leq p_q$, with $\rho = 0.5$ denoting the correlation between consecutive covariates $x_j$ and $x_{j+1}$.  We consider the following linear predictors 
\begin{align*}
    \Phi^{-1}\left( p^{1(1)}_{i} \right) = \eta^{(1)}_{i1}\ &= -1x_{i2} + 0.5x_{i3} + 1x_{i4}  -0.5x_{i6}
    , \\
    \log\left(\left(-\log(1 - p^{1(2)}_{i}\right) \right) = \eta^{(2)}_{i1}\  &=  0.5x_{i1} - 1x_{i2} + 0.75x_{i3} ,
     \\
    \tanh\left( \vartheta^{(c)}_i \right)^{-1} = \eta^{(c)}_{i} &=   0.5x_{i2} - 1.5x_{i3} +  1.5x_{i4}. 
\end{align*}

\begin{figure}[ht]
    \centering
      \includegraphics[scale = 0.65]{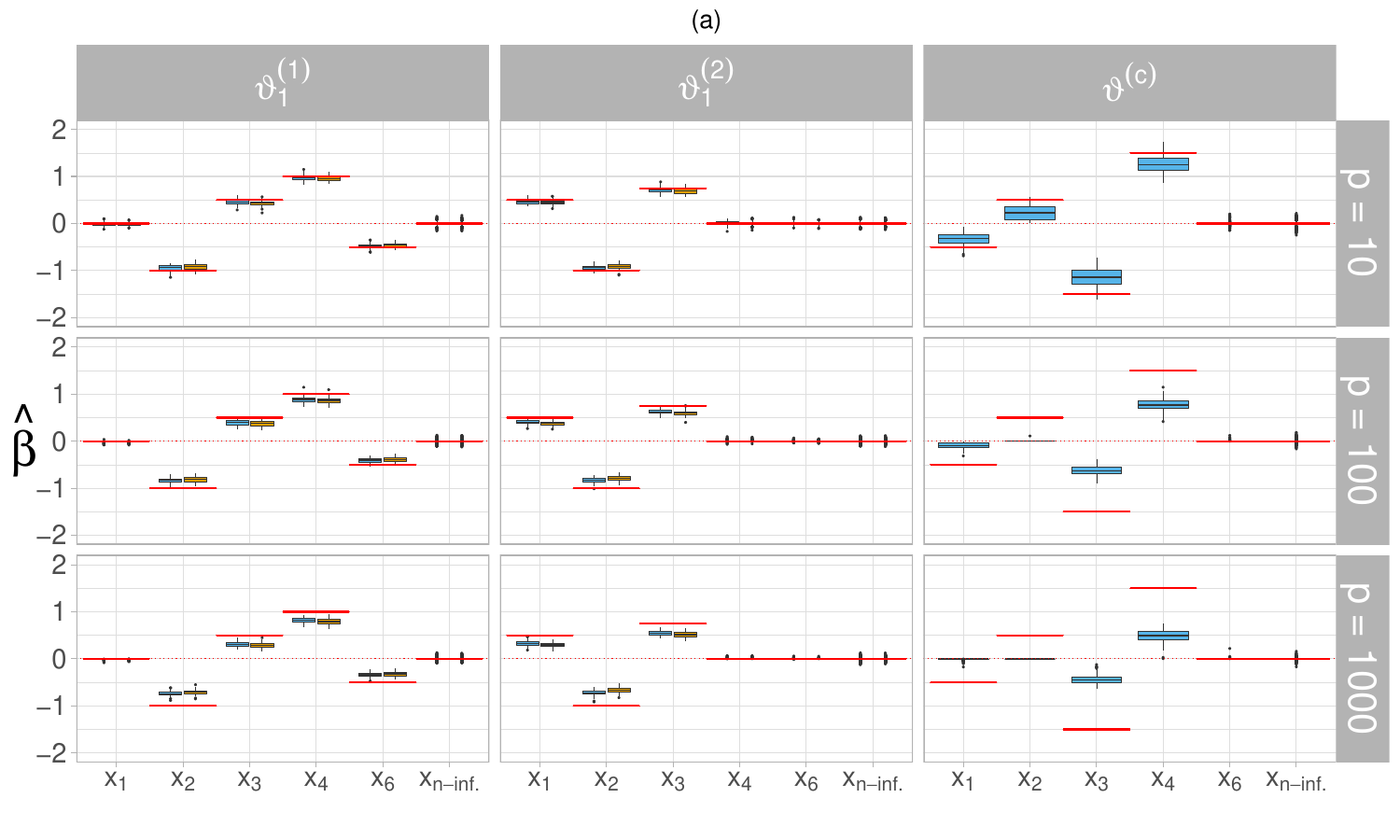}\\
    \includegraphics[scale = 0.65]{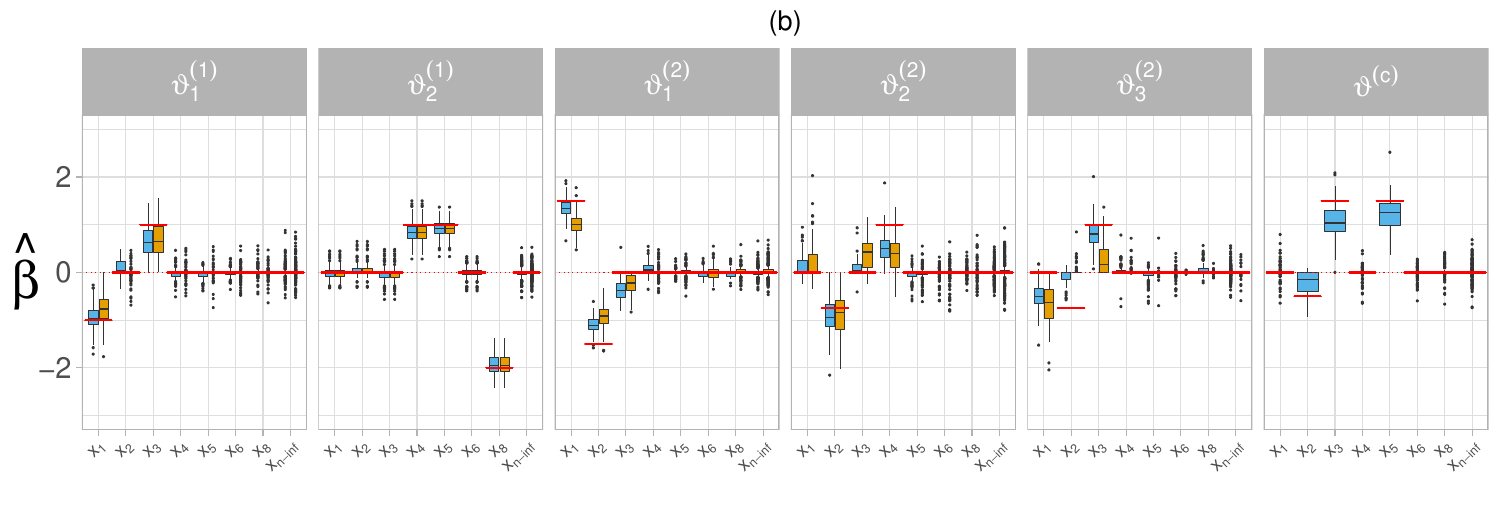} \\
 \includegraphics[scale = 0.65]{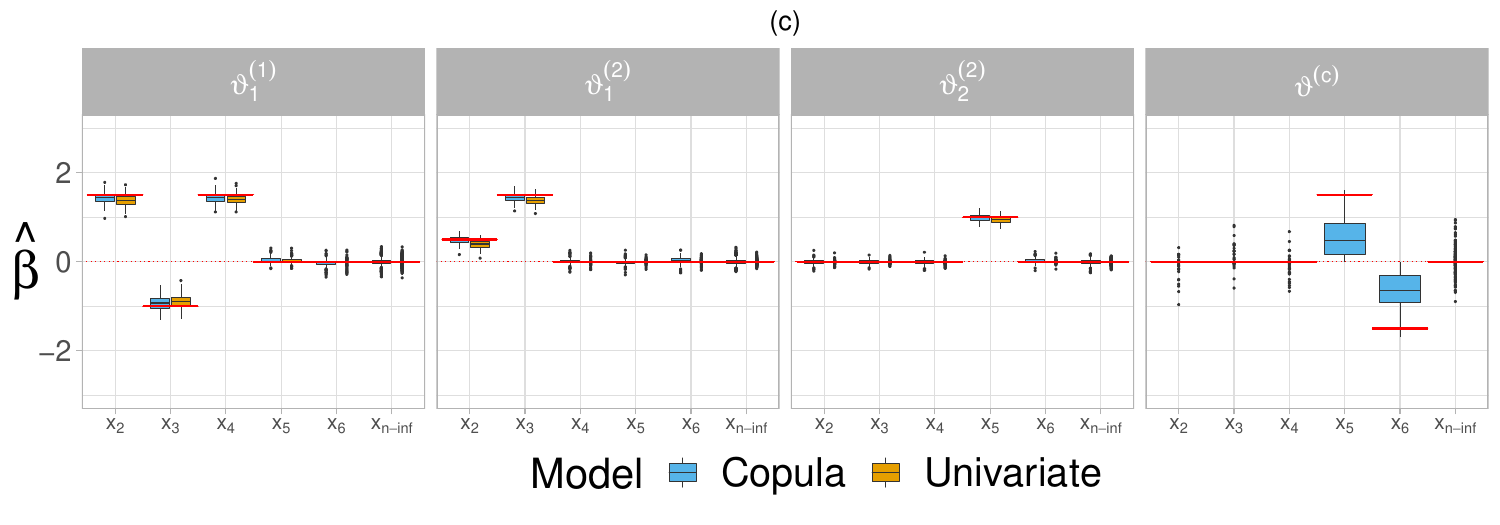} 
    \caption{\small Simulation study. Estimated coefficients of informative and non-informative ($x_{\text{n-inf.} }$) covariates from copula and univariate models across distribution parameters for linear DGPs of Simulation~\ref{sec:s1} (a, Gaussian copula), \ref{sec:s2} (b, Joe copula), \ref{sec:s3} (c, rotated Clayton copula by 270\degree ). %Red lines indicate true effects. 
    }
    \label{SIMULATIONSBOXPLOTS}
\end{figure}

\paragraph{Results}
Table~\ref{SimulationsAllMetrics}, Column~$(1)$ summarizes the performance scores. Overall, the copula model exhibits a better performance in terms of the negative log-likelihood (log-score) as well as the energy score, indicating a better  fit of the bivariate distribution. In terms of univariate scores (Brier score and AUC), both the univariate and copula models show similar results, with the copula model outperforming the univariate model at predicting the second margin in the high-dimensional setting ($p_3 = 1000$). The selection rates of informative and non-informative covariates corresponding to the individual distribution parameters are given in Table~\ref{SimulationsSelectionRates}, Column~(1). Across the board, the copula model tends to have a slightly higher selection rate of non-informative covariates than the univariate models, although these false positives decrease considerably as the number of non-informative covariates increases. The selection rates in the dependence parameter $\vartheta^{(c)}$ of non-informative regressors are considerably lower than that of effects in the marginal parameters, whereas that of informative covariates is slightly lower than 100\%. This suggests that the shrinkage is  strongest in the dependence parameter and the chosen criterion to determine $\texttt{m}_{\texttt{stop}}$ slightly underfits the effects in the dependence. However, both the copula and univariate models correctly select the informative covariates in all margins. Figure~\ref{SIMULATIONSBOXPLOTS}(a) depicts the estimated coefficients with each row corresponding to $p_1 = 10$, $p_2 = 100$, $p_3 = 1000$, respectively. In low-dimensional settings ($p_1$), the copula model matches the univariate models in producing accurate estimates of all linear coefficients in the marginal distributions. In high-dimensional settings ($p_3$), the estimated coefficients corresponding to the margins exhibit similar performance as in settings with $p_1$ and $p_2$.

\subsection{Bivariate discrete responses}\label{sec:s2}
\paragraph{Data generation}
We consider two DGPs for bivariate count responses each including $p = 10$ covariates. In the first DGP, the covariates have a strictly linear effect on the distribution parameters, whereas in the second DGP we consider non-linear effects. The bivariate discrete distribution is constructed using a combination of a Zero-Altered Logarithmic distribution (ZALG, margin~1) with two parameters, and a Zero-Inflated Negative Binomial Type I distribution (ZINBI, margin~2), which has three parameters. The marginal distributions and number of covariates are chosen to resemble the data studied in Section~\ref{SectionApplicationBivDiscrete}. The components are linked through a Joe copula, which allows to model positive dependence as well as upper tail dependence between the margins. The additive predictor of the dependence parameter $\vartheta_i^{(c)}$ covers Kendall's $\tau$ values  within $[0.275; \ 0.899]$, ranging from moderate to very strong positive dependence between $Y_1$ and $Y_2$. The covariates are sampled from independent univariate Uniform distributions with support between 0 and 1, i.e.\ $X_r \sim U[0,1], \ \forall r = 1, \dots, 10$. Overall, the bivariate distribution consists of six parameters with the following additive predictors. 

{\footnotesize
\begin{align*}
\text{Linear DGP:} & & \text{Non-linear DGP:}\\
   \log\left( \frac{\vartheta_{i1}^{(1)}}{1 - \vartheta_{i1}^{(1)}} \right) &=   - 1x_{i1}  + 1x_{i3}, & 
   \log\left( \frac{\vartheta_{i1}^{(1)}}{1 - \vartheta_{i1}^{(1)}} \right) &= \frac{1}{2}\left(x_{i1}^{3/2} - 2\cos(3x_{i1}) \right),  \\
   \log\left( \frac{\vartheta_{i2}^{(1)}}{1 - \vartheta_{i2}^{(1)}} \right) &=  + 1x_{i4} + 1x_{i5} - 2 x_{i8}, &
   \log\left( \frac{\vartheta_{i2}^{(1)}}{1 - \vartheta_{i2}^{(1)}} \right) &=  -80\left( x_{i3}^{3/2} - x_{i3}^{4/3} \right), \\ 
   \log\left( \vartheta_{i1}^{(2)} \right) &=  + 1.5 x_{i1} - 1.5 x_{i2}, & 
   \log\left( \vartheta_{i1}^{(2)} \right) &= -0.7\exp\left(x_{i2}^2 \right) + \exp\left( x_{i2}^{0.4} \right), %
\\  
\log\left( \vartheta_{i2}^{(2)} \right) &= -0.75 x_{i2}  + 1 x_{i4} ,  & 
\log\left( \vartheta_{i2}^{(2)} \right) &= 3 -1.5\left( 1.5\cos(2x_{i5}) + 3\tanh(x_{i5}) \right),  \\ 
\log\left( \frac{\vartheta_{i3}^{(2)}}{1 - \vartheta_{i3}^{(2)}} \right) &= -0.75 x_{i2} + 1 x_{i3}, & 
\log\left( \frac{\vartheta_{i3}^{(2)}}{1 - \vartheta_{i3}^{(2)}} \right) &= -3 -0.7\left( \sin(x_{i1}) - \exp(x_{i1})^2 \right), \\  
\log\left( \vartheta_i^{(c)} - 1\right) &= -0.5 x_{i2} + 1.5 x_{i3} + 1.5 x_{i5}, &
\log\left( \vartheta_i^{(c)} - 1\right) &= 2\sin(4x_{i4}).%
\end{align*}}

Note that in case of the linear DGP, seven out of the ten covariates have a non-zero effect on the distribution parameters with five of those overlapping and one having  an effect uniquely on one parameter. In the non-linear DGP, six covariates are informative and once again there is some overlap in the informative covariates across parameters. 

\paragraph{Results} The performance metrics for the bivariate count response scenario are summarized in Table~\ref{SimulationsAllMetrics}, Column $(2)$. In terms of  log- and energy scores, our proposed copula approach outperforms the univariate models considerably. The copula also leads to a smaller MSEP for predicting $Y_2$, whereas the univariate model for $Y_1$ outperforms the copula in terms of MSEP. The selection rates in Table~\ref{SimulationsSelectionRates}, Column (2) demonstrate that the copula model has a higher selection rate of non-informative regressors in the first margin, as well as in the first parameter of margin~2 compared to the univariate models in the linear DGP. However, the selection rates of informative covariates from the univariate models are considerably lower in the other two parameters of margin~2 compared to those of our copula model. The selection rates once again point out that the dependence parameter experiences the strongest shrinkage of the covariate effects. In the non-linear DGP, the selection rates of non-informative covariates are  similar for the copula and univariate models in the first margin. In contrast, in the second margin the copula model tends to select too many non-informative covariates in the first parameter of margin~2. Concerning the linear effects, our approach performs well given the overlap between covariate effects and distribution parameters. The shrinkage on the estimated coefficients corresponding to the dependence parameter exhibits a very similar behaviour to that observed in the bivariate binary scenario with $p_1 = 10$. The univariate models tend to underestimate the covariate effects in the second parameter of margin~1 and the third parameter of margin~2. Furthermore, the univariate models display a slightly higher variance in the estimated coefficients compared to those derived from the copula models. This observation is supported by Figure~\ref{SIMULATIONSFUNCTIONS}(b), which indicates that the selection rates of informative covariates in the non-linear DGP are consistently at 100\% across all models.

\subsection{Bivariate mixed responses}\label{sec:s3}
\paragraph{Data generation} For the mixed binary-continuous response scenario we generate the binary margin using a probit model, whereas the continuous margin follows a heteroskedastic Gaussian distribution. The components are linked through a Clayton copula rotated by 270$\degree$, which allows for dependence between very high values of $Y_1$ and low values of $Y_2$. The choice of margins and copula is based on the data on children malnutrition analysed in Subsection~\ref{SectionApplicationMixedBinCont}. A total of $p = 10$ covariates are obtained from independent univariate Uniform distributions between 0 and 1, i.e.\ $X_r \sim U[0,1], \ \forall r = 1, \dots, 10$. Once again we study both a DGP with only linear effects and another with non-linear effects of the covariates. The bivariate distribution features four parameters with following additive predictors.

\begin{footnotesize}
\begin{align*}
\text{Linear DGP:} & & \text{Non-linear DGP:}\\
  \Phi^{-1}\left( \vartheta_{i1}^{(1)} \right) &=  1.5 x_{i2} - 1x_{i3} + 1.5 x_{i4}, & 
   \Phi^{-1}\left( \vartheta_{i1}^{(1)} \right) &= \frac{1}{2}\left(x_{i1}^{3/2} - 2\cos(3x_{i1}) \right), \\  
  \vartheta_{i1}^{(2)} &= 0.5 x_{i2} + 1.5 x_{i3} , & 
   \vartheta_{i1}^{(2)}  &= -0.7\exp\left(x_{i1}^2 \right) + \exp\left( x_{i1}^{0.4} \right),\\ 
   \log\left(  \vartheta_{i2}^{(2)} \right) &=   1 x_{i5}, &  
    \log\left(  \vartheta_{i2}^{(2)} \right) &= -0.5 + \cos(2 x_{i2} )\\
    \log\left(  -\vartheta_{i}^{(c)} \right) &= 1.5 x_{i5} - 1.5 x_{i6}, & 
     \log\left(  -\vartheta_{i}^{(c)} \right) &= -1+ 3\sin(4x_{i3}).
\end{align*} 
\end{footnotesize}

 In the linear DGP, only five covariates have a non-zero effect on the distribution parameters and once again there is some overlap between informative covariates and distribution parameters. In the non-linear DGP, there are four informative covariates with some overlap between parameters and informative covariates.
 
\begin{figure}[t!]
    \centering
    \includegraphics[scale = 0.65]{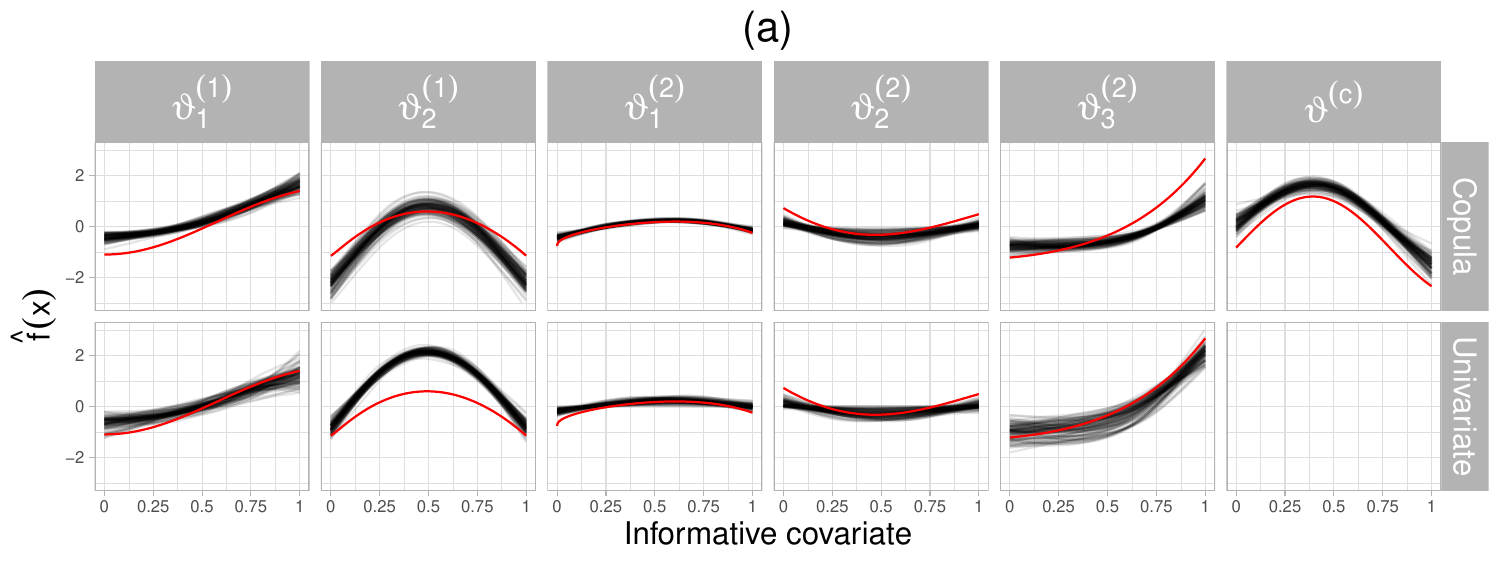} \\
 \includegraphics[scale = 0.65]{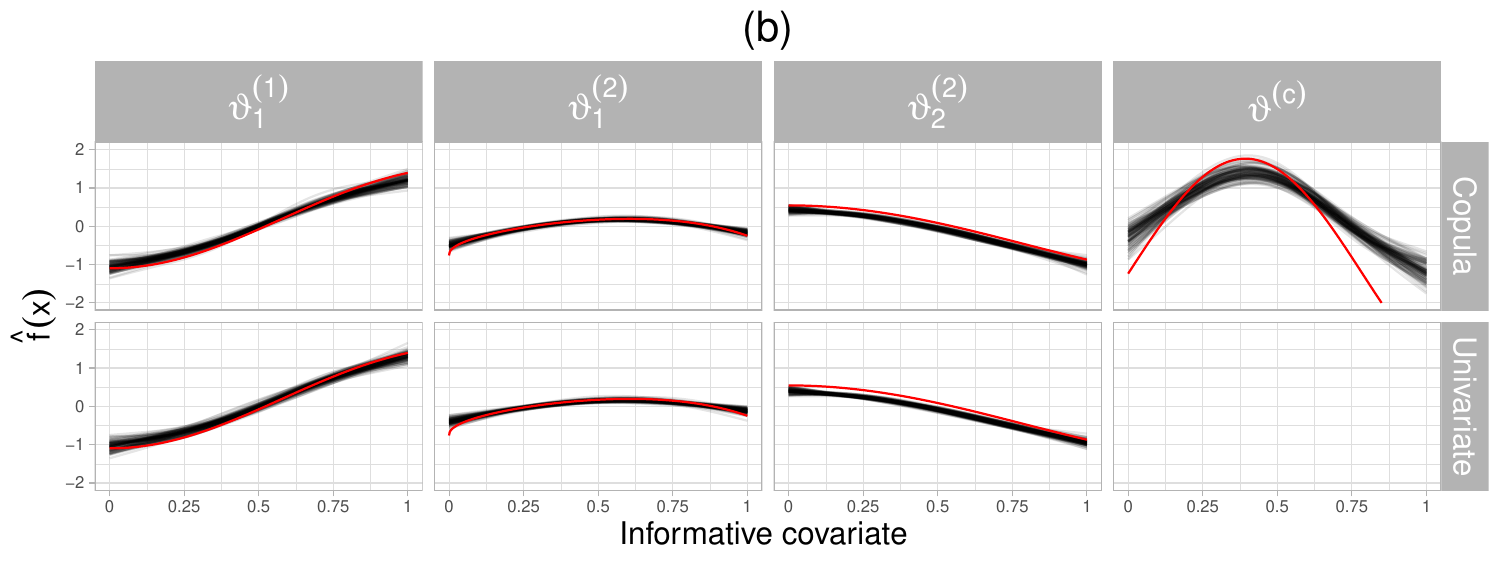} 
    \caption{\small Simulation study. Estimated effects of the informative covariates from copula and univariate models across distribution parameters for non-linear DGPs of Simulation~\ref{sec:s2} (a, Joe copula) and \ref{sec:s3} (b, rotated Clayton copula by 270\degree ). Red lines indicate true effects. 
    }
    \label{SIMULATIONSFUNCTIONS}
\end{figure}

\paragraph{Results} From Table~\ref{SimulationsAllMetrics}, Column~$(3)$ it can be observed that the copula model outperforms the univariate models in terms of the log and energy scores, with the difference in these two models becoming more apparent in the non-linear DGP. In the linear DGP, the copula model performs better than the univariate models in terms of the log-score but slightly worse in terms of the energy score, albeit the difference in energy scores is $0.001$ and the univariate models exhibit a much larger standard deviation in the aforementioned score. Regarding the univariate scores, once again both copula and univariate models exhibit similar performance. In the non-linear DGP, both models are able to recover the true non-linear effects of the informative covariates (see Figure~\ref{SIMULATIONSFUNCTIONS}(c)). In the linear DGP a slight improvement in efficiency can be observed from the copula model. We remark that the boosted copula model once again exhibits a higher degree of shrinkage of the effects present in the copula parameter, as indicated by the selection rates in  Table~\ref{SimulationsSelectionRates}, Column~(3). 
Similar to the other two response scenarios, both copula and univariate models effectively identify the informative covariates across all distribution parameters of the margins. In the non-linear DGP, the copula model demonstrates a tendency to exhibit higher selection rates of non-informative regressors within the continuous margin, while remaining competitive in the binary margin. 
\afterpage{\clearpage}

\begin{table}[ht]
    \centering
    \resizebox{\columnwidth}{!}{%
    \begin{tabular}{l l c c c c c c r  }
    \toprule
      &       & \multicolumn{3}{c}{$(1)$} & $(2)$ & $(3)$ \\
      \\
    Score  & Model & \multicolumn{3}{c}{Bivariate binary} & Bivariate count & Mixed \\
      &  &  $p_1 = 10$ &  $p_2 = 100$ & $p_3 = 1000$ & $p = 10$ & $p = 10$ \\
             %& \\
              %& \\  
         \midrule
             \rowcolor{Gainsboro!60}
     Log       & $C$ & $889.204 \ (21.187)$  & $914.826 \ (21.946)$ & $954.435 \ (21.371)$ & $1581.495 \ (48.501)$ & $1988.139 \ (28.164)$\\
               & $U$ & $958.433 \ (21.275)$  & $971.193 \ (22.645)$ & $997.890 \ (20.665)$ & $1963.842 \ (64.589)$ & $2007.379 \ (26.322)$ \\
                               \\
            \rowcolor{Gainsboro!60}
                               & $C\star$ & - & - & - & $2446.896 \ (58.728)$ & $1751.614 \ (32.803)$\\
                               & $U\star$ & - & - & - & $2828.331 \ (68.807)$ & $1906.526 \ (32.008)$ \\
      %                           & \\
     \midrule
     \rowcolor{Gainsboro!60}
     Energy              & $C$ & $0.280 \ (0.007)$ & $0.284 \ (0.007)$ & $0.293 \ (0.007)$ & $0.726 \ (0.040)$ & $0.686 \ (0.013)$ \\
                         & $U$ & $0.286 \ (0.007)$ & $0.289 \ (0.007)$ & $0.298 \ (0.007)$ & $0.742 \ (0.040)$ & $0.685 \ (0.505)$ \\
                               \\
     \rowcolor{Gainsboro!60}
                               & $C\star$ & - & - & - & $1.580 \ (0.074)$  & $0.669 \ (0.016)$ \\
                               & $U\star$ & - & - & . & $1.599 \ (0.073)$ & $0.676 \ (0.016)$ \\
     %& \\
      \midrule
      \rowcolor{Gainsboro!60}
     Brier ($Y_{1}$)       & $C$ & $0.142 \ (0.006)$ & $0.143 \ (0.007)$ & $0.148 \ (0.007)$ &  - & $0.199 \ (0.006)$ \\
                         & $U$ & $0.142 \ (0.006)$ & $0.143 \ (0.007)$ & $0.148 \ (0.006)$ &  - & $0.199 \ (0.006)$\\
                               \\
    \rowcolor{Gainsboro!60}
                               & $C\star$ & - & - & - & - &  $0.174 \ (0.006)$ \\
                               & $U\star$ & - & - & - & - &  $0.174 \ (0.006)$\\
                               \midrule
    \rowcolor{Gainsboro!60}
     Brier ($Y_{2}$)             & $C$ & $0.177 \ (0.006)$ & $0.180 \ (0.006)$ & $0.185 \ (0.006)$ &  - & - \\
                               & $U$ & $0.177 \ (0.006)$ & $0.180 \ (0.005)$ & $0.185 \ (0.006)$ &  - & - \\
                               \\
    \rowcolor{Gainsboro!60}
                               & $C\star$ & - & -  & -   & - & - \\
                               & $U\star$ & - & -  & - & - & - \\
     % & \\
      \midrule
      \rowcolor{Gainsboro!60}
     AUC ($Y_{1}$)               & $C$ & $0.879 \ (0.010)$ & $0.878 \ (0.012)$ & $0.870 \ (0.013)$ &  - & $0.762 \ (0.015)$\\
                               & $U$ & $0.878 \ (0.009)$ & $0.878 \ (0.012)$ & $0.871 \ (0.013)$ &  - & $0.762 \ (0.015)$\\
                               \\
    \rowcolor{Gainsboro!60}
                               & $C\star$ & - & - & - & - &  $0.816 \ (0.012)$\\
                               & $U\star$ & - & - & - & - &  $0.816 \ (0.012)$\\
                               \midrule
    \rowcolor{Gainsboro!60}
     AUC ($Y_{2}$)               & $C$ & $0.795 \ (0.015)$ & $0.789 \ (0.014)$ & $0.777 \ (0.016)$ &  - & -  \\
                               & $U$ & $0.795 \ (0.015)$ & $0.790 \ (0.014)$ & $0.781 \ (0.015)$ &  - & - \\
                               \\
    \rowcolor{Gainsboro!60}
                               & $C\star$ & - & - & -  & - & -  \\
                               & $U\star$ & - & - & -  & - & - \\
     %& \\
      \midrule
      \rowcolor{Gainsboro!60}
     MSEP ($Y_{1}$)              & $C$ & - & - & - & $1.082 \ (0.129)$ & - \\
                               & $U$ & - & - & - & $1.071 \ (0.129)$  & - \\
                               \\
    \rowcolor{Gainsboro!60}
                               & $C\star$ & - & - &  - & $1.567 \ (0.299)$  & - \\
                               & $U\star$ & - & - &  - & $1.560 \ (0.298)$ & - \\
                               \midrule
    \rowcolor{Gainsboro!60}
     MSEP ($Y_{2}$)              & $C$ & - & - & - &  $2.536 \ (0.512)$  & $1.189 \ (0.057)$ \\
                               & $U$ & - & - & - &  $2.443 \ (0.540)$ & $1.188 \ (0.057)$\\
                               \\
    \rowcolor{Gainsboro!60}
                               & $C\star$ & - & - & - & $10.727 \ (1.133)$  & $1.278 \ (0.075)$ \\
                               & $U\star$ & - & - & - & $10.971 \ (1.189)$  & $1.276 \ (0.074)$\\
%                               \midrule
%                               \midrule
%                               \rowcolor{Gainsboro!60}
%    Selection rate & $C$ & & & & & \\
%    $x_\text{inf}\ ;\ x_\text{n-inf.}$ & $U$ & & & & & \\
%     \\
%        \rowcolor{Gainsboro!60}
%     & $C\star$ & & & & & \\
%     & $U\star$ & & & & & \\
      \midrule
      \midrule
                                %\rowcolor{Gainsboro!60}
    Copula                 &  & \multicolumn{3}{c}{Gaussian} & Joe & Rotated Clayton 270\degree \\
     %\multicolumn{2}{l}{Kendall's $\tau$ mean} & & & & & & \\
    \multicolumn{2}{l}{Kendall's $\tau$ range}   & \multicolumn{3}{c}{ $[-0.993; \ 0.993]$ } & $[0.275; \  0.899]$ & \textbf{$[-0.787; \ -0.019]$} \\
    \\
    \multicolumn{7}{l}{Gradients stabilised using $L_2$ norm, step-length $\texttt{s}_{\texttt{step}} = 0.1$. $n_{\text{train}} = 1000$, $n_{\text{test}} = 1000$, $n_{\texttt{mstop}} = 1500$. } \\
\bottomrule
    \end{tabular}
    }
    \caption{\small Simulation study. Performance metrics for the simulation studies for the copula ($C$) and univariate models ($U$), $\star$ identifies the non-linear DGP. Values are mean scores from the 100 independent replicates (each evaluated on  the test dataset), whereas
    parentheses show the respective standard deviations. }
    \label{SimulationsAllMetrics}
\end{table}

\afterpage{\clearpage}

\begin{table}[ht]
    \centering
    \resizebox{\columnwidth}{!}{%
    \begin{tabular}{l cccccccc c c c c c c c }
    \toprule
           &   & \multicolumn{8}{c}{$(1)$} & & \multicolumn{2}{c}{ $(2)$ } & &  \multicolumn{2}{c}{ $(3)$ } \\
      \\
  & & \multicolumn{8}{c}{Bivariate binary} & &  \multicolumn{2}{c}{ Bivariate count } & & \multicolumn{2}{c}{ Binary \& continuous } \\
        & & \multicolumn{2}{c}{ $p_1 = 10$ } & &  \multicolumn{2}{c}{ $p_2 = 100$ } & & \multicolumn{2}{c}{ $p_3 = 1000$ } & & \multicolumn{2}{c}{ $p = 10$ } & & \multicolumn{2}{c}{ $p = 10$ } \\
             %& \\
   & &  $x_{\text{inf}}$ & $x_{\text{n-inf}}$ & & $x_{\text{inf}}$ & $x_{\text{n-inf}}$ & & $x_{\text{inf}}$ & $x_{\text{n-inf}}$
   & & $x_{\text{inf}}$ & $x_{\text{n-inf}}$  & & $x_{\text{inf}}$ & $x_{\text{n-inf}}$ \\
              %& \\  
         \midrule
         %  \multirow{15}{*}{\STAB{\rotatebox[origin=c]{90}{rota}}} \\
         \multicolumn{16}{c}{Linear DGP}  \\
        % \midrule
   \multicolumn{16}{l}{Copula model $(C)$ }  \\
   \\
  \rowcolor{Gainsboro!60}
$\vartheta_{1}^{(1)}$ & & $100$ & $77.800$ &  & $100$ & $25.500$ & & $100$ & $6.093$  & & $99$ & $63.250$ &  & $100$ & $63.571$ \\
$\vartheta_{2}^{(1)}$ & & -   & -  &  & - & - & & - & - & & $100$ & $66.143$ &  & - & - \\
\\
 \rowcolor{Gainsboro!60}
$\vartheta_{1}^{(2)}$ & & $100$ & $78.600$ & & $100$ & $27.900$ & & $100$ & $7.088$ & & $100$ & $74.875$ &  & $100$ & $72.625$ \\
$\vartheta_{2}^{(2)}$ & & -  & - & & - & - & & - & - & & $98$ & $52.750$ &  & $100$ & $82.556$ \\
 \rowcolor{Gainsboro!60}
$\vartheta_{3}^{(2)}$ & & - & - & & - & - & & - & - & & $76.500$ & $49.875$ &  & - & - \\
\\
$\vartheta_{}^{(c)}$ & & $96.500$ & $57.500$ & & $71.000$ & $6.400$ & & $56.250$ & $0.221$ & & $90$ & $42.286$ &  & $87.500$ & $28.125$ \\
\midrule
 \multicolumn{16}{l}{Univariate models $(U)$ }  \\ %%  UNIVARIATE MODEL!!!
 \\
 \rowcolor{Gainsboro!60}
$\vartheta_{1}^{(1)}$ & & $100$ & $69.800$ & & $100$ & $21.900$ & & $100$ & $4.778$ & & $97$ & $39.125$ &  & $100$ & $49.857$ \\
$\vartheta_{2}^{(1)}$ & & - & - & & - & - & & - & - & & $100$ & $46$ &  & - & - \\
\\
 \rowcolor{Gainsboro!60}
$\vartheta_{1}^{(2)}$ & & $100$ & $70.700$ & & $100$ & $20.700$ & & $100$ & $4.594$ & & $100$ & $60.250$ &  & $100$ & $30.875$ \\
$\vartheta_{2}^{(2)}$ & & - & - & & - & - & & - & - & & $91.500$ & $57.125$ &  & $100$ & $37.667$ \\
 \rowcolor{Gainsboro!60}
$\vartheta_{3}^{(2)}$ & & - & - & & - & - & & - & - & & $38$ & $18.625$ &  & - & - \\
\midrule
\midrule
\multicolumn{16}{c}{Non-linear DGP}  \\
%%%%%%%%% NON-LINEAR SETTINGS
%\midrule
 \multicolumn{16}{l}{Copula model $(C\star)$ }  \\
 \\
  \rowcolor{Gainsboro!60}
$\vartheta_{1}^{(1)}$ & & - & - & & - & - & & - & - &  & $100$  & $30.556$ &  & $100$ & $35.444$ \\
$\vartheta_{2}^{(1)}$  & & - & - & & - & - & & - & - & & $100$  & $22.222$  &  & - & - \\
\\
 \rowcolor{Gainsboro!60}
$\vartheta_{1}^{(2)}$ & & - & - & & - & - & & - & - & & $100$ & $92.444$ &  & $100$ & $82.667$ \\
$\vartheta_{2}^{(2)}$ & & - & - & & - & - & & - & - & & $100$ & $33.444$ &  & $100$ & $84.333$ \\
 \rowcolor{Gainsboro!60}
$\vartheta_{3}^{(2)}$ & & - & - & & - & - & & - & - & & $100$ & $4.889$ &  & - & -\\
\\
$\vartheta_{}^{(c)}$ & & - & - & & - & - & & - & - & & $100$ & $19.556$ &  & $100$ & $0.444$ \\
\midrule
 \multicolumn{16}{l}{Univariate models $(U\star)$ }  \\%%  UNIVARIATE MODEL!!!
 \\
 \rowcolor{Gainsboro!60}
$\vartheta_{1}^{(1)}$ & & - & - & & - & - & & - & - & & $100$ & $27.000$ &  & $100$ & $38.444$ \\
$\vartheta_{2}^{(1)}$ & & - & - & & - & - & & - & - & & $100$ & $22.222$ &  & - & - \\
\\
 \rowcolor{Gainsboro!60}
$\vartheta_{1}^{(2)}$ & & - & - & & - & - & & - & - & & $99$ & $61.333$ &  & $100$ & $24.778$ \\
$\vartheta_{2}^{(2)}$ & & - & - & & - & - & & - & - & & $98$ & $53.556$ &  & $100$ & $36.444$ \\
 \rowcolor{Gainsboro!60}
$\vartheta_{3}^{(2)}$ & & - & - & & - & - & & - & - & & $100$ & $8.778$ &  & - & -\\
\bottomrule
    \end{tabular}
    }
    \caption{\small Simulation study. Selection rates (in \%) of informative ($x_{\text{inf}}$) and non-informative covariates ($x_{\text{n-inf.}}$) for the copula ($C$) and univariate models ($U$) for each distribution parameter, $\star$ denotes non-linear DGP. Values are averages over the $100$ independent datasets. 
    }
    \label{SimulationsSelectionRates}
\end{table}

\afterpage{\clearpage}

\afterpage{\clearpage}

\subsection{Overall summary of simulation results} 

In general, the performance of the proposed boosted copula models is satisfactory. They effectively detect and recover all effects across different parameters of the bivariate distribution. Notably, the copula dependence parameter shows stronger shrinkage of informative effects compared to other parameters. As the number of considered covariates increases, the degree of shrinkage also rises. This behaviour may be attributed to the greedy nature of the algorithm, since a reduction of the loss from including a covariate with a small coefficient in the dependence parameter might not be large enough compared to updating a coefficient in any other parameter corresponding to the margins. Consequently, this can lead to sparser dependence parameters, potentially resulting in certain informative covariates with relatively smaller effects being falsely disregarded.

The choice of $\texttt{m}_{\texttt{stop}}$ in distributional copula models remains an under-explored area, deserving attention in future research to address this issue. Overall, the copula approach is competitive in terms of selection rates of covariates in the marginal parameters and satisfactory in identifying the most relevant effects in the dependence parameter. It clearly demonstrates benefits when it comes to evaluating the predictive behaviour in terms of probabilistic scores across a wide range of marginal distributions and dependence structures.  This highlights the added value compared to using boosting with independent univariate models.

\section{Biomedical Applications}\label{SECTIONCaseStudy}

In this section we illustrate the versatility of our proposed boosted distributional copula regression approach by analysing three different biomedical research questions. In Section~\ref{SectionApplicationBivBinary} we model the joint distribution of two binary responses which correspond to the presence of heart disease (yes/no) as well as the presence of high cholesterol (yes/no) using data from the large-scale biomedical database UK Biobank \citep{bycroft2018uk}. This corresponds to a high-dimensional setting in the covariate space. In Section~\ref{SectionApplicationBivDiscrete} we are concerned with the joint distribution of a bivariate count vector comprised of the number of doctor consultations and the number of prescribed medications from Australian healthcare recipients using data from the \texttt{R} package \texttt{bivpois} \citep{KarlisBivPoisson}. We demonstrate how to conduct model-building by means of the predictive risk when the choice of marginal distributions as well as copula function is not clear. Lastly, in Section~\ref{SectionApplicationMixedBinCont} we investigate the distribution of two mixed responses relevant for analysing infant malnutrition in India emanating  using data from the Demographic and Health Survey (DHS, \url{https://dhsprogram.com}). In what follows, the step-length of the boosting algorithm is set to $\texttt{s}_{\texttt{step}} = 0.1$ and the number of fitting iterations $\texttt{m}_{\texttt{stop}}$ is optimised via the predictive or out-of-bag risk as outlined in Step (5) of Algorithm~\ref{boostingAlgorithm}. We resort to $L_2$-stabilisation in order to achieve similar effective step-lengths across the different parameters of the bivariate distributions. Additionally, we let all covariates of each respective application to potentially enter all parameters of the bivariate distribution at the beginning of the fitting procedure, allowing the boosting algorithm to select the relevant covariates in each parameter in a data-driven manner. 

\subsection{Chronic ischemic heart disease and high cholesterol}\label{SectionApplicationBivBinary}

We analyse the joint distribution of high cholesterol (yes/no) and chronic ischemic heart disease (yes/no) using a sample from the large-scale UK Biobank genetic cohort study \citep{bycroft2018uk} under application number 81202. We analyse a subsample consisting of $n = 30{,}000$ individuals and $p = 1{,}867$ pre-filtered genetic variants (covariates). This sample has been previously analysed in \cite{StroemerDistReg} using  a bivariate Bernoulli distribution. The prevalence of the two factors in our sample is 7.2\% and 32.3\%, respectively. 
\paragraph{Model specification} In contrast, we construct the joint distribution using a Gaussian copula with logit margins. In our case the dependence structure is given by the correlation coefficient  $\vartheta^{(c)}$, whereas the bivariate Bernoulli distribution uses the odds ratio. Our copula approach has the advantage that we obtain a directly interpretable dependence measure (Pearson's correlation) with $\hat{\vartheta}^{(c)}$. Additionally, it is also possible to compute the dependence in terms of Kendall's $\tau$ as well as the odds ratio once the copula has been estimated. Therefore our method is more general than the bivariate Bernoulli distribution and offers complete flexibility in terms of the dependence structure as well as choice of the link functions used to model the margins. We split the sample into two partitions dedicated for fitting ($n_{\text{train}} = 20{,}000$) and tuning of $\texttt{m}_{\texttt{stop}}$ ($n_{\texttt{mstop}} = 10{,}000$). The additive predictors of the bivariate distribution are
\begin{equation*}
   \eta_{i1}^{(\bullet)}  = \beta_{01}^{(\bullet)} +  \sum_{r = 1}^{ 1,867 } \beta_{ r1 }^{(\bullet)}  x_{ir},\  \text{ with } \bullet = \{1,2,c\}.
\end{equation*}

%\afterpage{\clearpage}
\begin{figure}[t]
    \centering
    \includegraphics[scale=0.45]{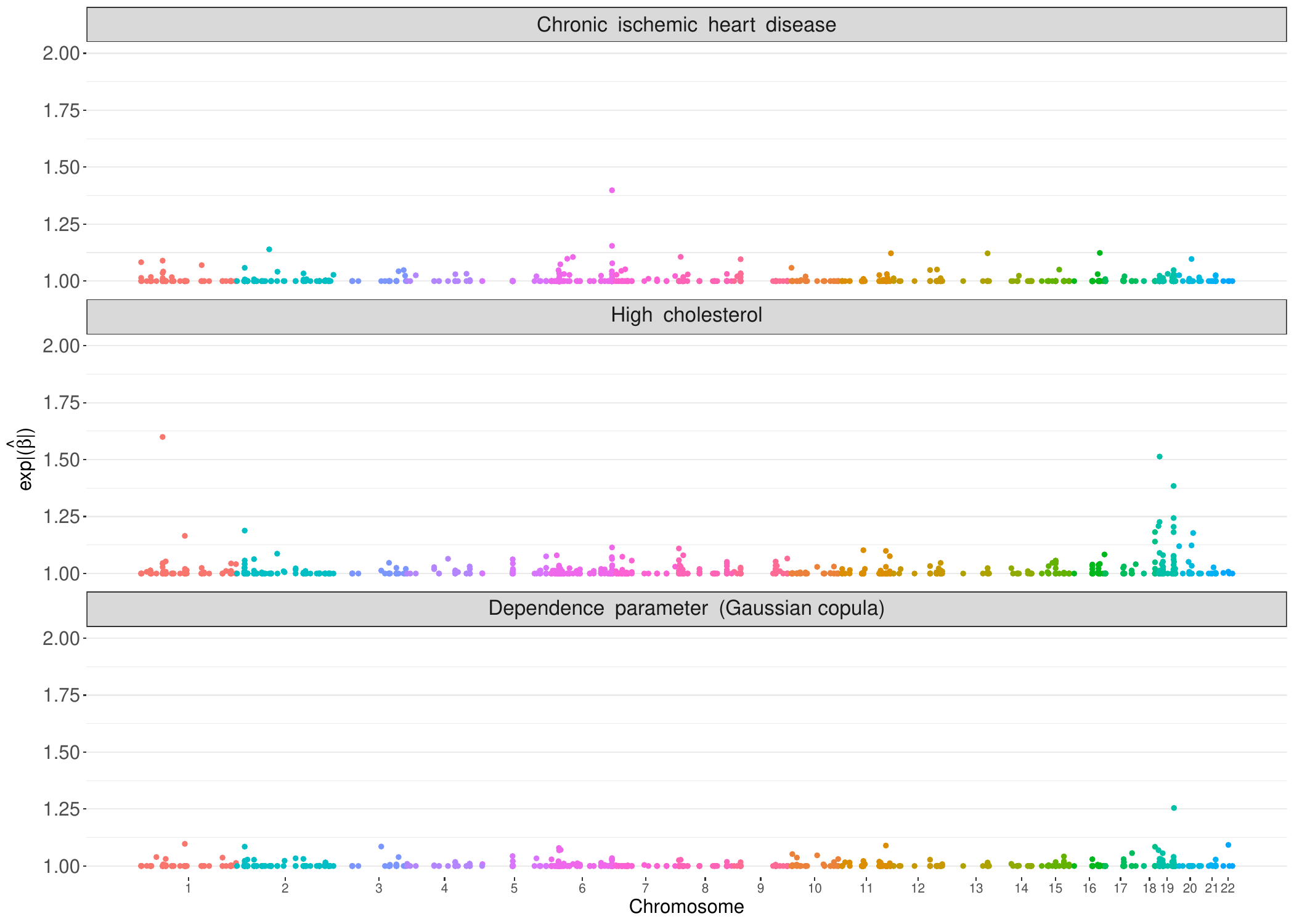}
    \caption{\small Application~\ref{SectionApplicationBivBinary}. Manhattan-type plots of the estimated coefficients (expressed in exponential absolute values of the estimated values) of the boosted bivariate binary model using a Gaussian copula. The $x$-axis represents the genomic location of the variants. }
    \label{FigureApplicationBivBin}
\end{figure}

\paragraph{Results} The estimated coefficients expressed as exponential absolute values in each margin and the dependence parameter are shown in Figure~\ref{FigureApplicationBivBin}.  The fitted dependence values in the sample expressed in terms of Kendall's $\tau$ are within $\hat{\tau} \in [-0.567;\  0.289]$. This result indicates that there is a strong negative dependence between the probabilities of chronic heart disease and high cholesterol. This finding most likely reflects the common use of statins in the population of patients already diagnosed with chronic heart disease \citep{sinnott2021genetics}.  Our proposed boosting method selects several variants in the respective parameters of the bivariate distribution. For instance, out of a potential $1{,}867$ possible candidates, 140 variants are selected in the first margin ($\vartheta_{1}^{(1)}$), 322 variants in the second margin ($\vartheta_{1}^{(2)}$) and 181 in the dependence parameter $\vartheta_{}^{(c)}$ with some overlap in the selected variants between the parameters (90 variants selected for two out of three parameters). A total of 19 variants are shared between the dependence parameter and $\vartheta_{1}^{(1)}$, whereas $\vartheta_{1}^{(2)}$ and $\vartheta_{}^{(c)}$ have 48 variants in common. Moreover, 23 variants are shared among the margins. The findings of our copula model agree with previous studies on the location of cholesterol-associated genes \citep[see e.g,][]{Richardson2020}, where the highest estimated coefficient values are present. We remark that shrinkage on the dependence parameter of our approach is less pronounced than that observed in the analysis of \cite{StroemerDistReg}, hence our model is able to detect more variants that alter the dependence between high cholesterol and ischemic heart disease.

\subsection{Doctor consultations and prescribed medications in Australia}\label{SectionApplicationBivDiscrete}
We study the joint distribution of a bivariate count response comprised of the number of doctor consultations ($\texttt{doctorco} \in \mathbb{N}$) and the number of prescribed medications ($\texttt{prescrib} \in \mathbb{N}$) of healthcare recipients from Australia. The sample consists of $n=5{,}190$ observations and we use $75\%$ of them to fit the model ($n_{\text{train}} = 3{,}892$), and the remaining 25\% for optimising $\texttt{m}_\texttt{stop}$ ($n_{\text{mstop}} = 1{,}298$). The dataset comprises two continuous covariates. These are   \texttt{age} (age in years divided by 100) and \texttt{income} (annual income in Australian dollars divided by 1000). In addition, the binary covariate  \texttt{gender} (1 female, 0 male) is reported.

\paragraph{Marginal distributions} A preliminary screening of univariate distributions on the individual margins was conducted with the best-fitting distribution being determined by means of the predictive risk. As shown in Figure~\ref{FIGUREThreeScatterplots}(b), each of the marginal responses exhibit a large amount of zeros and their respective variances differ from the mean ($\overline{\texttt{doctorco}} = 0.302$; $Var(\texttt{doctorco}) = 0.637$, and $\overline{\texttt{prescrib}} = 0.863$; $Var(\texttt{prescrib}) = 2.003$). In line with these descriptive statistics, we find that the Poisson distribution is not suited to model the conditional distribution of the two responses. The best-fitting marginal distributions in terms of predictive risk are the Zero-Altered Logarithmic distribution $\left( \vartheta_{1}^{(1)}, \vartheta_{2}^{(1)} \right)$ for \texttt{doctorco}, in this case the expectation and variance are determined by both parameters. The Zero-Inflated Negative Binomial distribution $\left(\vartheta_{1}^{(2)}, \vartheta_{2}^{(2)}, \vartheta_{3}^{(2)} \right)$ is the most suitable for \texttt{prescrib}, the parameters $\vartheta_{1}^{(2)}$ and $\vartheta_{2}^{(2)}$ determine the mean, whereas all three parameters determine the variance, see Table~\ref{TableMarginalDistributions}. Moreover, the probability of observing a zero is explicitly modelled via $\vartheta_{3}^{(2)}$. 
\paragraph{Copula selection}
The copula was selected by means of the predictive risk out of a number of five possible candidates (Gaussian, Frank, Clayton, Gumbel, FGM and AMH copulas) with the Clayton copula giving the best predictive risk.  This indicates that the data supports the presence of lower tail dependence, i.e.\ strong dependence of very low values in both marginal responses. 

\paragraph{Predictor specification}
As a result of the selection of marginal distributions, there are six parameters in the bivariate distribution ($K_1 = 2,\, K_2 = 3,\, K_c = 1$) and all additive predictors in the distribution share the following configuration.
\begin{equation*}
   \eta_{ik}^{(\bullet)}  = \beta_{0k}^{(\bullet)} + \beta^{(\bullet)}_{1k} \texttt{gender}_i + s^{(\bullet)}_{1k}\left( \texttt{income}_i \right) + s^{(\bullet)}_{2k}\left( \texttt{age}_i \right), \quad \forall k = 1, \dots, K,\ \bullet \in \{1,2,c\}.
   \end{equation*}
We use P-spline base-learners for the covariates \texttt{income} and \texttt{age} as well as a linear base-learner for the covariate \texttt{gender}.

\begin{figure}[ht]
    \centering
    \includegraphics[scale = 0.65]{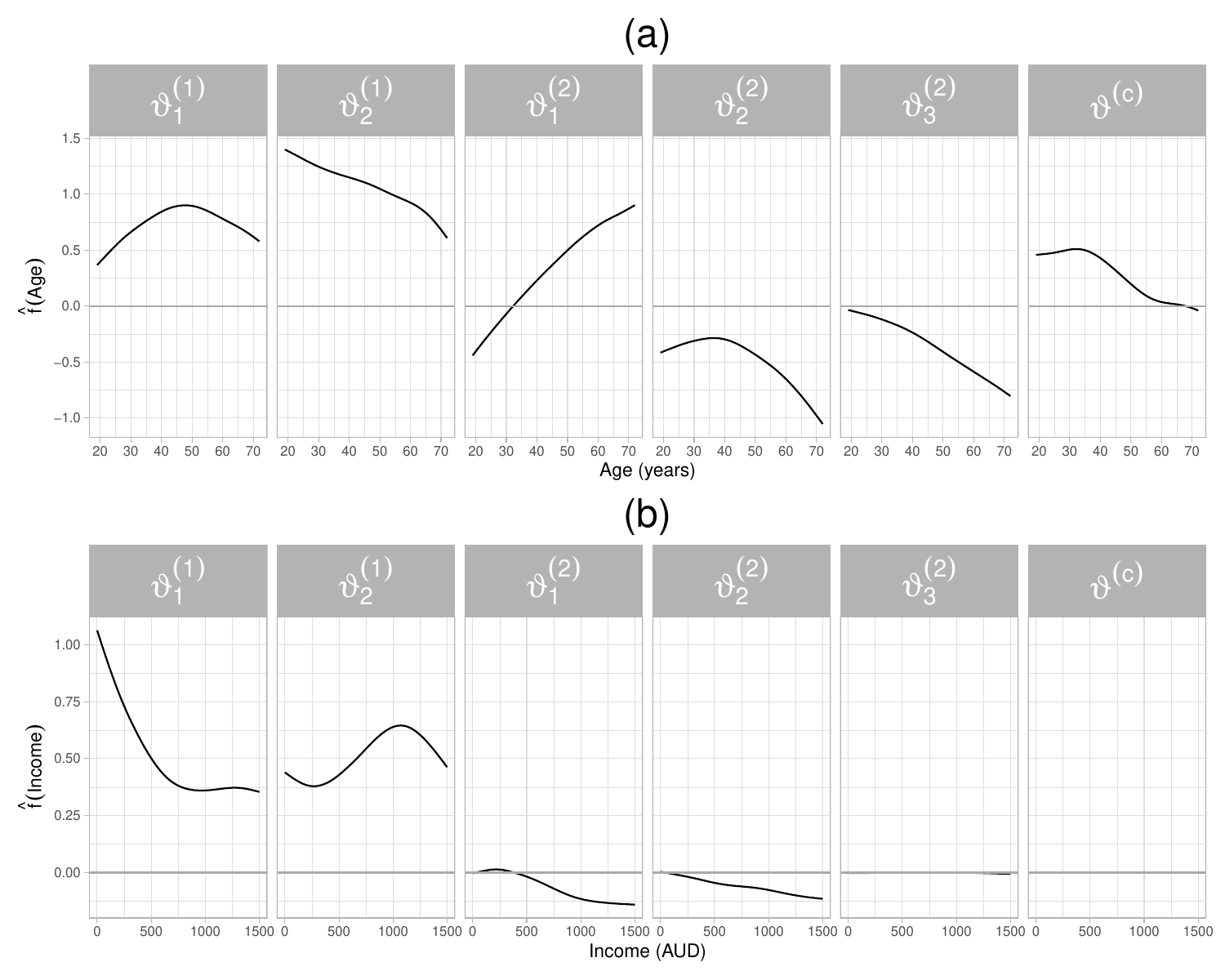}
    \caption{\small Application \ref{SectionApplicationBivDiscrete}. Estimated non-linear effects of \texttt{age} and \texttt{income} on the parameters of the margins as well as the dependence parameter of a Clayton copula.}
    \label{APPLICATIONDISCRETE}
\end{figure}

\paragraph{Results} The fitted values of the dependence expressed as Kendall's $\tau$ range within $\hat{\tau} \in [0.336;\ 0.564]$, indicating a moderate to strong estimated dependence between the margins in the sample.  The results of non-linear effect estimates are depicted in Figure~\ref{APPLICATIONDISCRETE}. The covariate \texttt{age} appears to have a non-zero effect in all parameters of the bivariate distribution (see Figure~\ref{APPLICATIONDISCRETE} (a)). In particular,  the effect of age on $\vartheta_{1}^{(1)}$ is increasing between 20 and 50 years, and then becomes decreasing for older individuals. On the other hand, age leads to smaller values of the parameter $\vartheta_{2}^{(1)}$. These two parameters jointly determine the expectation and variance of \texttt{doctorco}. The individual's age leads to an increase on the predictor of $\vartheta_{1}^{(2)}$, which partially determines the expected number of prescribed medications. Intuitively, the predictor of $\vartheta_{3}^{(2)}$ decreases almost linearly with the individual's age, which directly translates to the logit of a decreased probability of observing a zero in the second margin. In other words older individuals are expected to get more likely a non-zero number of prescribed medications. A downward-sloping effect of age is also estimated for the parameter $\vartheta_{2}^{(2)}$. Additionally, the dependence between the margins decreases in older individuals as seen in the panel corresponding to $\vartheta^{(c)}$. The covariate \texttt{income} (in Australian dollars, AUD) is selected in four parameters of the bivariate distribution, see Figure~\ref{APPLICATIONDISCRETE} (b). The individual's income has a non-zero effect on the parameters of \texttt{doctorco} distribution. Conversely, \texttt{income} exhibits a much smaller, albeit downward-sloping, effect on the parameters $\vartheta_{1}^{(2)}$ and $\vartheta_{2}^{(2)}$ of the distribution of \texttt{prescrib}. The covariate \texttt{income} was not selected on the dependence parameter and its effect on $\vartheta_{3}^{(2)}$ is very close to zero. The covariate \texttt{gender} was selected in all parameters except for $\vartheta_{1}^{(1)}$, compare Table~\ref{APPLICATION_COEFFICIENTSTABLE}, middle block. The estimates of \texttt{gender} in the first margin indicate that expected value of both responses is higher for female healthcare recipients, \textit{ceteris paribus}.  The estimated effect of \texttt{gender} in $\vartheta_{3}^{(2)}$ also suggests that the probability of having zero prescribed medications is lower for female recipients compared to male individuals. Lastly, the dependence between the margins is lower for female individuals, relative to their male counterparts. 

\begin{table}[ht]
    \centering
     \resizebox{\columnwidth}{!}{%
    \begin{tabular}{ll c rr c rrr c c}
    \toprule
%                 & & \multicolumn{6}{c}{Parameter  } \\
              &   & & \multicolumn{2}{c}{margin~1 } & & \multicolumn{3}{c}{margin~2 } & & Copula  \\
 Application    &  Covariate & & $\vartheta_{1}^{(1)}$ & $\vartheta_{2}^{(1)}$ & & $\vartheta_{1}^{(2)}$ &  $\vartheta_{2}^{(2)}$ & $\vartheta_{3}^{(2)}$ & & $\vartheta_{}^{(c)}$\\
       \midrule
          \multirow{1}{*}{Bivariate binary}   & & & \multicolumn{2}{c}{Bernoulli (logit)} & & \multicolumn{3}{c}{Bernoulli (logit)} & & \multicolumn{1}{c}{Gaussian}\\
      \\
      \rowcolor{Gainsboro!60}
           & \texttt{Intercept} & & $-1.198$ & -- & & $-0.317$ & -- & -- & & $\phantom{-}0.442$ \\
       %  & & & & & & &  & &  \\
    \midrule
    \midrule
      \multirow{1}{*}{Bivariate count}   & & & \multicolumn{2}{c}{ZALG} & & \multicolumn{3}{c}{ZINBI} & & \multicolumn{1}{c}{Clayton}\\
      \\
      \rowcolor{Gainsboro!60}
           & \texttt{Intercept} & & $-1.198$ & $-0.049$ & & $-0.317$ & $0.234$  & $0.030$ & & $\phantom{-}0.442$ \\
      & \texttt{gender} (female) &  & 0 & $-0.228$ & & $0.275$ & $-0.644$ & $-1.171$ & & $-0.390$ \\
       %  & & & & & & &  & &  \\
    \midrule
    \midrule
       \multirow{1}{*}{Bivariate mixed}   & & & \multicolumn{2}{c}{Bernoulli (probit)} & & \multicolumn{3}{c}{Gaussian} & & \multicolumn{1}{c}{Clayton $270\degree$ }\\
       \\
    \rowcolor{Gainsboro!60}
   & \texttt{Intercept}    & & $-0.230$ & -- & & $0.003$ & $0.008$ & -- & & 0 \\
    & \texttt{cgender} (female) & & $-0.031$ & -- & & 0 & $0.002$ & -- & &  0 \\
    % & \texttt{breastfeeding} & &  & - & & & & - & 0 \\
    %      & & & & & & & & & \\
         \bottomrule
    \end{tabular} }
    \caption{\small Estimated linear effects for Applications  \ref{SectionApplicationBivBinary} (first block), \ref{SectionApplicationBivDiscrete} (second block)  and \ref{SectionApplicationMixedBinCont} (third block) across distribution parameters. The symbol ``--'' indicates that the distribution does not feature the respective parameter, whereas $0$ indicates that the algorithm did not select the respective covariate. }
    \label{APPLICATION_COEFFICIENTSTABLE}
\end{table}

%\afterpage{\clearpage}

\subsection{Determinants of infant malnutrition in India }\label{SectionApplicationMixedBinCont}

We analyse a sample of $n = 24{,}286$ observations to study jointly two determinants of child malnutrition in India.  The binary response $\texttt{fever} \in \{0,1\}$ indicates whether a child has had fever up to two weeks prior to the survey interview, whereas $\texttt{wasting} \in \mathbb{R}$ denotes low weight-for-height, indicating an acute recent weight loss. According to UNICEF, this is the most immediate, visible and life-threatening form of malnutrition \citep{UNICEFWASTING}. The individuals in the sample are spread across 438 administrative units (districts) with some imbalance in the number of observations per district. We resort to a slightly different sub-sampling scheme compared to the previous applications in order to obtain $n_{\text{train}}$, and $n_{\texttt{mstop}}$: We include all observations corresponding to districts with a sample size below or equal to 40 in $n_{\text{train}}$. For all other district with more than 40 observations, we sample without replacement and obtain a fraction of around 75\% of the total observations used for training ($n_{\text{train}} = 18{,}214$) and 25\% for optimizing \texttt{mstop} ($n_{\texttt{mstop}} = 6{,}072$). Table~\ref{TableApplicationVariableDescriptions3} summarizes responses and available covariates. 

\begin{table}[ht]
    \centering
       \resizebox{\columnwidth}{!}{%
    \begin{tabular}{l l lc c}
    \toprule
       Variable  & Description & Type &  & Mean (s.d.) \\
       \midrule
        \texttt{fever} & Fever experienced within two weeks preceding survey interview & Binary & & $0.307 \ (0.461)$\\
        \texttt{wasting}  & Low weight-for-height & Continuous & & $-79.144 \ (123.367)$ \\
           \midrule
         \texttt{cage} & Age of the child in months & Continuous & & $17.255 \ (10.148) $ \\
          \texttt{breastfeeding} & Months of breastfeeding & Continuous & & $14.076 \ (8.751)$\\
           \texttt{mbmi} & Mother's Body-Mass-Index & Continuous & & $19.783 \ (2.937)$ \\
           \midrule
                       \texttt{cgender} & Gender of the child (1 female, 0 male) & Binary & & $0.476 $\\
            \texttt{distH} & District of residence & Factor & & - \\
            % \texttt{} & & & & \\
          \midrule
           \multicolumn{5}{l}{Number of districts: $438$, $n = 24{,}286$.} \\
         \bottomrule
    \end{tabular} }
    \caption{\small Variables of Section~\ref{SectionApplicationMixedBinCont}. Responses are fever (binary, row 1) and wasting (continuous, row 2);  covariates entering the model non-linearly are age of child, breastfeeding and the mother's body-mass-index (rows 3--5); the binary covariate gender of child enters the model linearly (row 5) and the district in India (row 6) enters the model as a discrete spatial effect.}
    \label{TableApplicationVariableDescriptions3}
\end{table}

\paragraph{Model specification} We follow \cite{KleinBinaryContinuous} and set the link function for the model of \texttt{fever} to probit, whereas for \texttt{wasting} we resort to a heteroskedastic Gaussian distribution. The dependence between the margins is modelled using a Clayton copula rotated by $270\degree$. This allows to model dependence between very high values of \texttt{fever} and very low values of \texttt{wasting}. It seems reasonable to expect such a dependence structure being supported by the data, since it is likely that the probability of children experiencing fever is prone to be dependent with low weight-for-height values (\texttt{wasting}, i.e.\ undernourished infants). Consequently, the bivariate distribution has $K=4$ distribution parameters. In \cite{KleinBinaryContinuous} the additive predictor of the margins was fixed and an information-criterion-based model selection procedure was conducted using different configurations of the predictor of $\vartheta^{(c)}$. Here we allow our proposed approach to select the variables in all predictors of the bivariate distribution in a data-driven manner, without further input from the analyst. That is,

{\footnotesize
\begin{align*}
   \eta_{ik}^{(\bullet)}  &= \beta_{0k}^{(\bullet)} + \beta^{(\bullet)}_{1k} \texttt{cgender}_i + s^{(\bullet)}_{1k}\left( \texttt{cage}_i \right) + s^{(\bullet)}_{2k}\left( \texttt{mbmi}_i \right) + %\\
   %&\phantom{=} 
   s^{(\bullet)}_{3k}\left( \texttt{breastfeeding}_i \right) + s^{(\bullet)}_{4k}\left( \texttt{distH}_i \right),
\end{align*}}
where $k = 1, \dots, K,\ \bullet \in \{1,2,c\}$ and $s^{(\bullet)}_{4k}\left( \texttt{distH}_i \right)$ is set as a Markov Random Field base-learner to model the discrete spatial information of the districts in the data. The covariates \texttt{cage}, \texttt{mbmi} and \texttt{breastfeeding} are incorporated using  P-spline base-learners with 20 knots and second order difference penalties, whereas a linear base-learner is used for \texttt{cgender}. 

\begin{figure}[ht]
    \centering
    \includegraphics[scale = 0.62]{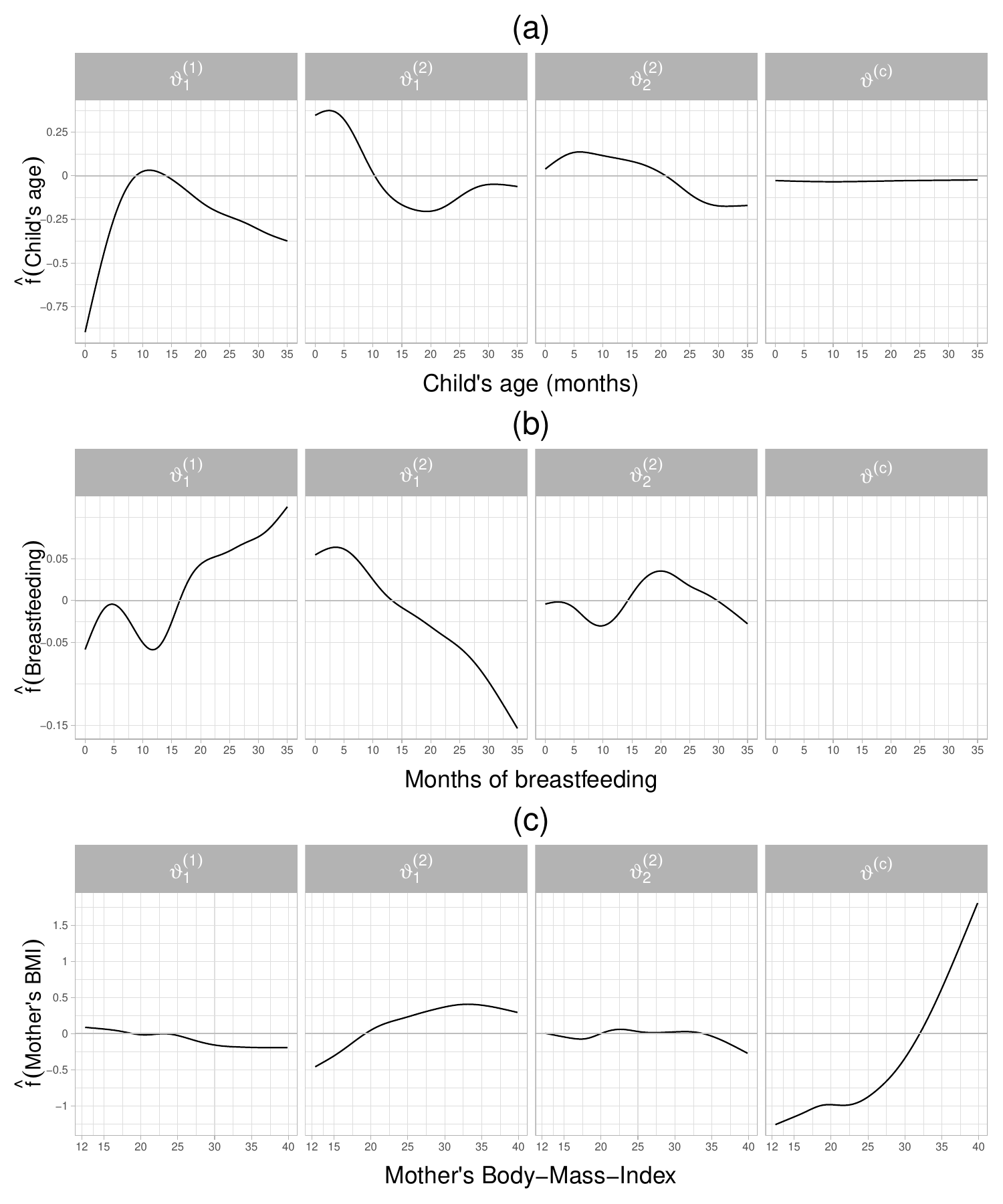}
    \caption{\small Application \ref{SectionApplicationMixedBinCont}. Estimated non-linear effects of \texttt{cage}, \texttt{breastfeeding} and \texttt{mbmi} on the parameters of the margins as well as the dependence parameter of a Clayton copula rotated by $270\degree$. }
    \label{APPLICATIONMIXED_ALLCONTINUOUSCOVARIATES}
\end{figure}

\begin{figure}[ht]
    \centering
    \includegraphics[scale = 0.62]{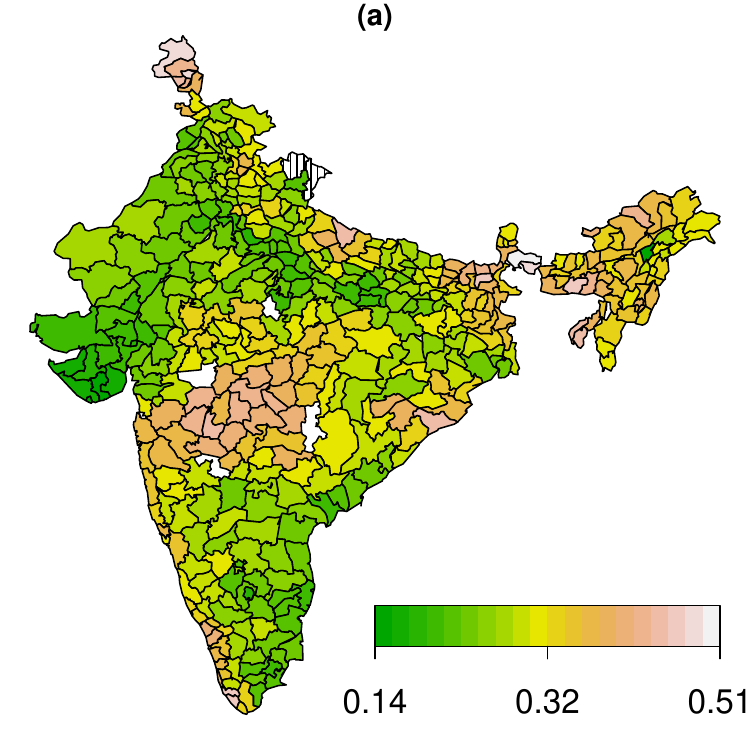}
    \includegraphics[scale = 0.62]{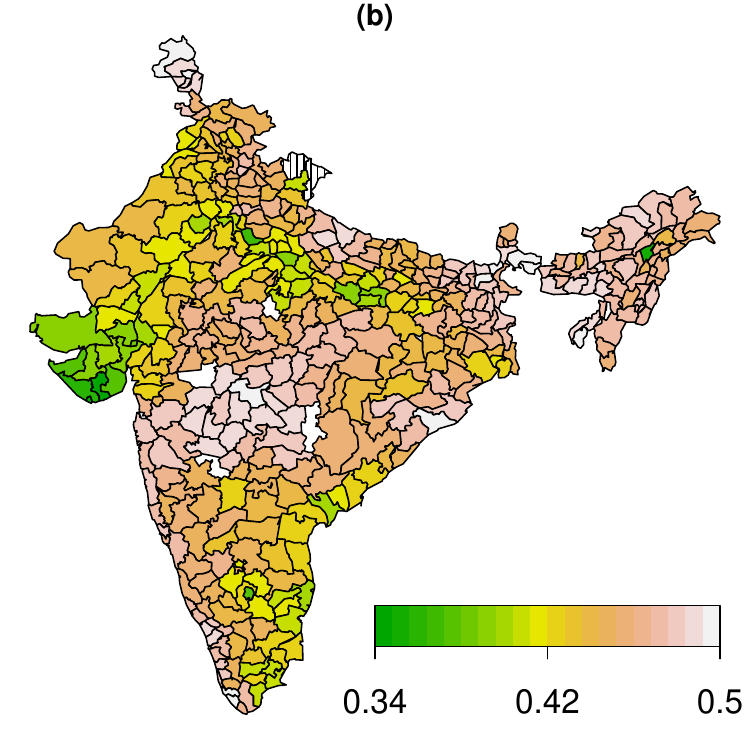}\\
      \includegraphics[scale = 0.62]{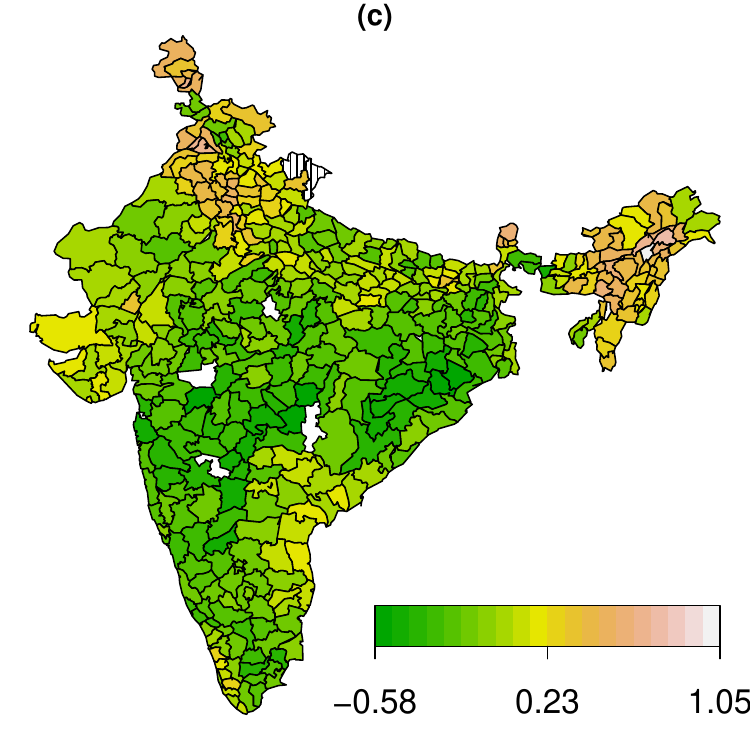}  
          \includegraphics[scale = 0.62]{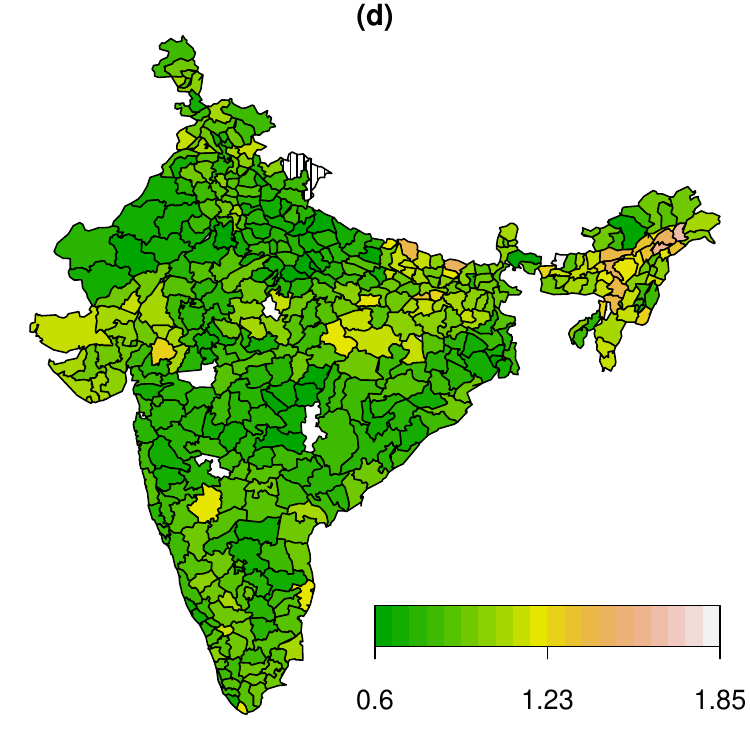} \\
      \includegraphics[scale = 0.62]{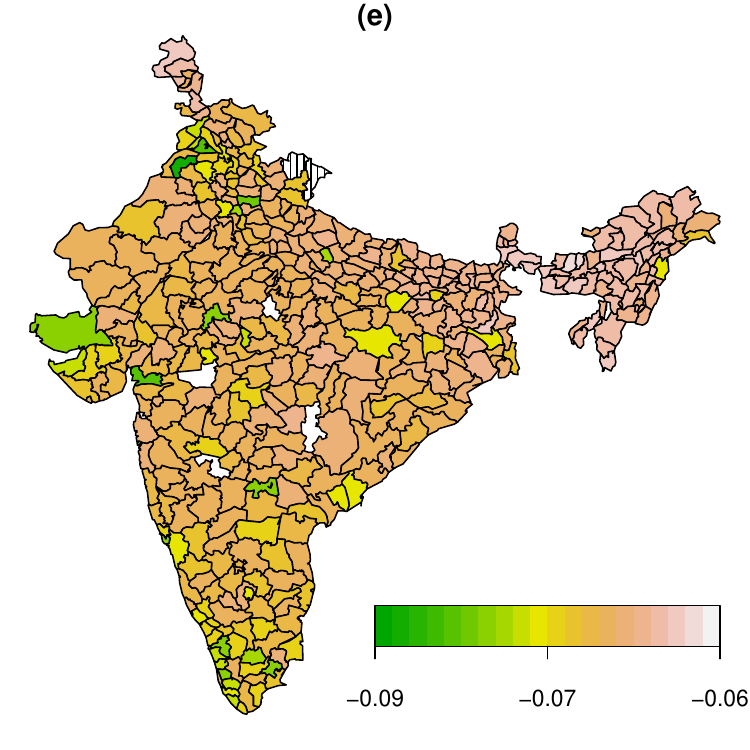}
     \includegraphics[scale = 0.62]{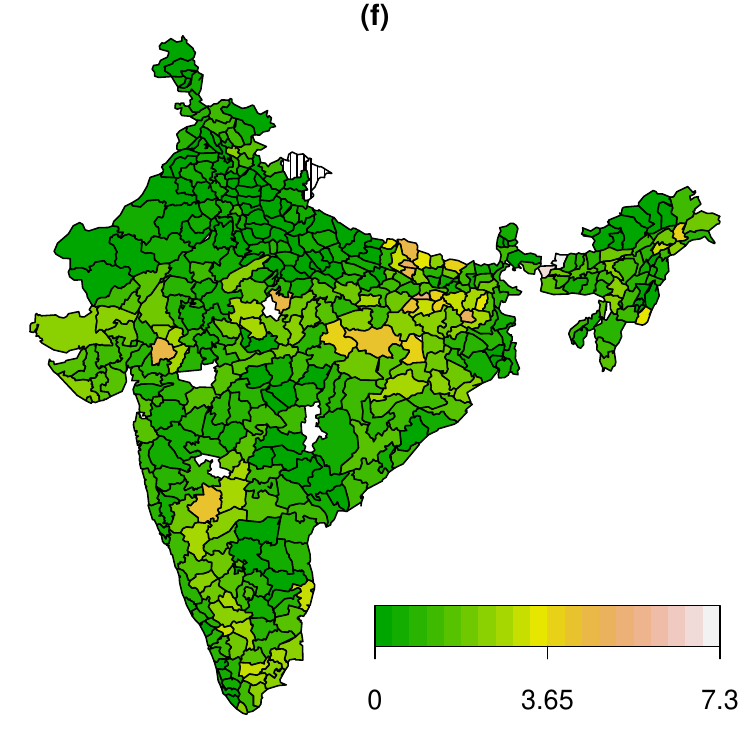}
    \caption{Application \ref{SectionApplicationMixedBinCont}. Shown are (a) expected value, and (b) standard deviation  of  \texttt{fever}; (c) estimated expected value,  and (d) standard deviation  of \texttt{wasting}; (e) estimated Kendall's $\tau$,  and (f) joint probabilities in \% of having fever and moderate undernutrition  according the Clayton copula rotated by $270\degree$.}
    \label{APPLICATIONMIXED_SPATIALGRID}
\end{figure}

\paragraph{Results} The estimated dependence between the margins in terms of Kendall's $\tau$ ranges within $\hat{\tau} \in [-0.561;\ -0.052]$, suggesting a negative dependence between \texttt{wasting} and \texttt{fever}. This is a reasonable finding since a lower \texttt{wasting} score implies a more severe form of undernutrition, whereas the risk of fever is expected to be positively associated with a poor health status. The estimated non-linear effects of the covariates \texttt{cage}, \texttt{breastfeeding} and \texttt{mbmi} are visualized in Figure~\ref{APPLICATIONMIXED_ALLCONTINUOUSCOVARIATES}.  It can be seen that children within 0 and $\approx 12$ months of age have an increasing likelihood of \texttt{fever}. The estimated effect of \texttt{cage} is downward-sloping in the first twenty months on the expectation of \texttt{wasting}, whereas on the standard deviation a similar pattern is observed albeit with a much smaller slope. In terms of the dependence structure, the child's age appears to have a negligible effect. The estimated effect of \texttt{breastfeeding} on \texttt{fever} shows an upward slope and on $\vartheta_{1}^{(1)}$ a downward slope. The presence of breastfeeding at a later age of the child could reflect a lack of other sources of nourishment apart from the mother, serving as a proxy for household's poverty, thus driving the probability of \texttt{fever} upwards and the expected value of \texttt{wasting} downwards. The variable \texttt{breastfeeding} is not selected in the dependence parameter. Compared to \texttt{cage} and \texttt{breastfeeding}, the mother's body-mass-index (\texttt{mbmi}) shows a small to moderate (see $\vartheta_{1}^{(2)}$) association with the margins. The effect of \texttt{mbmi} is slightly increasing in the expectation of \texttt{wasting} and remains stable at around $\texttt{mbmi} \approx 25$. However, the effect of \texttt{mbmi} leads to a sharp increase on the dependence between the margins after it reaches values of approximately 25. The covariate \texttt{cgender} was not selected in $\vartheta^{(2)}_{1}$ as well as $\vartheta^{(c)}_{}$ and it shows a very small value in $\vartheta_{2}^{(2)}$, compare Table~\ref{APPLICATION_COEFFICIENTSTABLE}, third block. Finally,  Figure~\ref{APPLICATIONMIXED_SPATIALGRID} presents various estimated quantities (expectation, standard deviation and Kendall's $\tau$, joint probabilities) according to the spatial structure of the data. The spatial component modelling the districts (\texttt{distH}) is selected in all parameters. In Figure~\ref{APPLICATIONMIXED_SPATIALGRID}(a) it can be observed that the districts located in the center of India exhibit higher probability of \texttt{fever}, however the standard deviation of \texttt{fever} is rather high across the country (see Figure~\ref{APPLICATIONMIXED_SPATIALGRID} (b)). The expectation of \texttt{wasting} remains mostly low throughout all districts, with some exceptions located in the north and north-eastern districts of India, see Figure~\ref{APPLICATIONMIXED_SPATIALGRID}(c). Compared to \texttt{fever}, the standard deviation of \texttt{wasting} is rather low in most districts, see Figure~\ref{APPLICATIONMIXED_SPATIALGRID} (d).  Figure~\ref{APPLICATIONMIXED_SPATIALGRID} (e--f) visualize  the per-district average of the estimated dependence between the margins in terms of Kendall's $\tau$ and the estimated joint probabilities (in \%) of having fever and moderate undernutrition i.e.,~$P(Y_1 = 1, Y_2 < -2)$. It can be seen that the magnitude of the dependence is larger in some districts located in the north-western are, as well as the south-eastern coast of India. The joint probabilities of fever and moderate undernutrition indicate that children located in mid-eastern districts are more prone to suffer from malnutrition. 
\afterpage{\clearpage}
 
\afterpage{\clearpage}
\afterpage{\clearpage}

\section{ Discussion }\label{SECTIONDiscussion}
We have extended the boosted distributional copula regression approach to accommodate arbitrary response types on different domains.  We conducted a wide range of simulation studies to investigate the predictive performance as well as the estimation capabilities of our proposed method. Overall, we found that our approach outperforms univariate boosting models when it comes to probabilistic forecasting. We remark that the shrinkage of covariate effects is stronger in the dependence parameter of the copula, indicating that the algorithm favours sparse, even sometimes independent margins before adopting a constant (intercept-only) or varying dependence structure.  

We were able to demonstrate that our proposed copula approach allows to capture the nuances of each marginal response, such as zero-inflation, over-dispersion, or heteroskedasticity, while also modelling the dependence between the margins using only one statistical model. Additionally, our methodology and software implementation allow to conduct data-driven variable selection without further input from the analyst as well as transparent and reproducible research. 

We have illustrated the application of our approach on three diverse  biomedical datasets and these analyses have been carried out using a portable computer (MacBook Pro with M1 Pro CPU and 32GB of memory), highlighting the efficiency and scalability of our software implementation. In the first application we identified relevant genetic variants associated with the dependence of high cholesterol  and ischemic heart disease. In contrast to previous analyses using a boosted distributional regression approach, the dependence structure of our Gaussian copula model is more interpretable and flexible. Additionally, the shrinkage on the dependence parameter is less pronounced compared to previous studies that used a boosted distributional regression model. Although not conducted here due to computation time constraints, our catalogue of implemented copula functions could be tested in order to investigate whether the data supports lower or upper tail dependence. In our second healthcare-related application we found that data on the number of doctor consultations and number of prescribed medications supports lower tail dependence i.e.\ dependence between extremely low values of the margins. Finally, in the third application we studied the joint distribution of two determinants of infant malnutrition that emanate from different domains. One determinant is expressed as a binary indicator whereas the other is a continuous marker.   

 The main limitation of resorting to statistical boosting for model fitting is the lack of confidence intervals for the estimated effects. These uncertainty quantification measures can be estimated using bootstrap methods, albeit it can be a cumbersome and time-consuming task \citep{hepp2019significance}. Another limitation was observed  in our simulation studies in Section~\ref{SECTIONSimStudy}: The boosted models have a tendency to select false positives throughout the fitting process and the different distribution parameters. Although the estimated effect of these false positives is in most cases small or negligible, a formal correction of these incorrectly estimated effects would be appealing. An adaptation of the de-selection procedure implemented by \cite{StroemerDESELECTION} would lead to more sparse models and stable selection of informative covariates.   
 
 Another future field of application  where data-driven variable selection can have a big impact is in observational studies where endogenous variables are present, see e.g.\ \cite{BrisenoSanchezFIVDR} and \cite{SAMPLESELECTIONCOUNTDATA}. Statistical boosting could provide valuable insights in these scenarios, since the effect of endogenous variables are identifiable as long as so-called instruments are available, which boosting could help identify and to validate the analyst's beliefs. Lastly, we are also exploring an extension of our boosting methodology to fit distributional copula regression models for bivariate time-to-event data, which would greatly extend the applicability of our software implementation in biomedical research.

\section*{Acknowledgements}
The work on this article was supported by the Deutsche Forschungsgemeinschaft (DFG, grant number 428239776, KL3037/2-1, MA7304/1-1). The authors gratefully acknowledge the computing time
provided on the Linux HPC cluster at Technical University
Dortmund (LiDO3), partially funded in the course of the
Large-Scale Equipment Initiative by the DFG via the
project 271512359. The analysis on the UK Biobank was conducted under the application number 81202.

\bibliographystyle{apalike}
\bibliography{Thebibliography}

\end{document}